\documentclass[preprint,eqsecnum,prd,nofootinbib,floatfix]{revtex4}
\usepackage{graphicx,amssymb}

\begin{document}

\title{Inclusive Jet Production, Parton Distributions,
\\
and the Search for New Physics}

\author{Daniel Stump}
\author{Joey Huston}
\author{Jon Pumplin}
\author{Wu-Ki Tung}
\affiliation{%
Department of Physics and Astronomy \\
Michigan State University \\
East Lansing, MI 48824}

\author{H.\,L.\ Lai}\affiliation{%
Department of Physics and Astronomy \\
Michigan State University \\
East Lansing, MI 48824}
\affiliation{%
Department of Science Education \\
Taipei Municipal Teachers College, Taiwan}

\author{Steve Kuhlmann}
\affiliation{%
Argonne National Laboratory \\
Argonne, IL 60439}

\author{J.\,F. Owens}
\affiliation{%
Department of Physics \\
Florida State University \\
Tallahassee, FL 32306}

\date{March 3, 2003}
\preprint{MSUHEP-030303}

\begin{abstract}
Jet production at the Tevatron probes some of the smallest distance
scales currently accessible.
A gluon distribution that is enhanced at large $x$
compared to previous determinations provides a better
description of the Run 1b jet data from both CDF and D\O.
However, considerable uncertainty still remains regarding
the gluon distribution at high $x$.
In this paper, we examine the effects of this uncertainty,
and of the remaining uncertainties in the NLO QCD theory,
on jet cross section comparisons to Run 1b data.
We also calculate the range of contributions still
possible from any new physics.
Predictions are also made for the expanded kinematic range
expected for the ongoing Run 2 at the Tevatron and for the LHC.
\end{abstract}

\maketitle

\clearpage
\tableofcontents
\clearpage

\setcounter{section}{0}

\section{Introduction: Jet cross sections at the Tevatron}
 
Jet production at the Tevatron probes the highest momentum transfer region 
currently accessible.
As this region is potentially sensitive to a wide variety of new
physics, there was great interest when the inclusive jet cross section
measured by the CDF collaboration in Run 1b exhibited an excess in
the highest $E_T$ range, when compared to NLO predictions using then-current
parton distribution functions~\cite{cdf1A}.
In an attempt to
determine if conventional physics could explain the deviation, the CTEQ PDF
fitting group carried out a global analysis using information from
deep-inelastic scattering and Drell-Yan data,
as well as from jet data from the Tevatron,
but giving a higher statistical emphasis to the
high $E_T$ jet data from CDF~\cite{cteq4hj}.
NLO predictions using the resulting fit
(CTEQ4HJ) reduced the size of the excess observed by CDF.
The jet data from both CDF and D\O\ were also used in a more conventional fit
(CTEQ4M) where no special emphasis was given to the
high $E_T$ data~\cite{cteq4}.

The  dominant subprocess that contributes to
jet production at high $E_T$ is quark-(anti)quark scattering.
However, the quark distributions in the corresponding $x$ range
are very well constrained by the precise DIS and DY data used
in the global fits.
Only the gluon distribution has the flexibility to change
significantly in the high $x$ region, and indeed the gluon
distribution increases by roughly a
factor of 2 for $x$ values of about 0.5 in the CTEQ4HJ fit.
The gluon-quark scattering subprocesses increase from
approximately 20\% of the total jet cross section at
high $E_T$ using CTEQ4M to 40\% using CTEQ4HJ~\cite{cteq4hj}.

The next group of PDF's from CTEQ (CTEQ5)~\cite{cteq5}
also contained two sets: CTEQ5M, the standard lowest $\chi^2$ solution, 
and CTEQ5HJ, defined by a similar statistical
enhancement applied to the high $E_T$ data from
CDF~\cite{cdf1b} and D\O\ ~\cite{d01b}.

More recently, D\O\ has measured the inclusive jet cross section as a
function of rapidity $y$ over the range $0\leq |y|\leq 3$ \cite{d0rapidity}.
This data set, comprised of 90 data points, has a greater statistical
power in the global fits than the CDF and D\O\ jet cross section
measurements in the central region alone.
The CTEQ6M fit\,\cite{cteq6} utilizes these D\O\ jet
cross section measurements, along with the CDF measurements
in the central rapidity region,
as well as the most recent DIS data from HERA
and existing fixed target DIS and DY data.
An important step was taken to quantify the uncertainties of
the parton distributions and their physical predictions with
the inclusion of 40 eigenvector basis PDF sets,
along with the best fit,
to characterize the uncertainties in the neighborhood of
the global minimum, using the Hessian method.

This paper is devoted to a detailed and more focused study
of the uncertainties of the inclusive jet cross section
due to global constraints, both at current energies,
and more importantly for further QCD studies and the
search for new physics at future collider programs.
During the course of this work,
we have produced an improved version of the CTEQ6 PDF's.
The new minimum set, called CTEQ6.1M henceforth, provides
a global fit that is almost equivalent in every respect
to the published CTEQ6M\ \cite{cteq6},
although some parton distributions
(e.g., the gluon) may deviate from CTEQ6M in some kinematic
ranges by amounts that are well within the specified uncertainties.
The more significant improvements are associated with some of the
40 eigenvector sets, which are made more symmetrical and reliable
in CTEQ6.1M\footnote{%
The CTEQ6.1M parton distributions will be available
in the CTEQ table format at www.phys.psu.edu/~cteq/
and in the LHAPDF format at vircol.fnal.gov.}

\begin{figure}[ht]
\begin{center}
\includegraphics[width=0.8\textwidth]{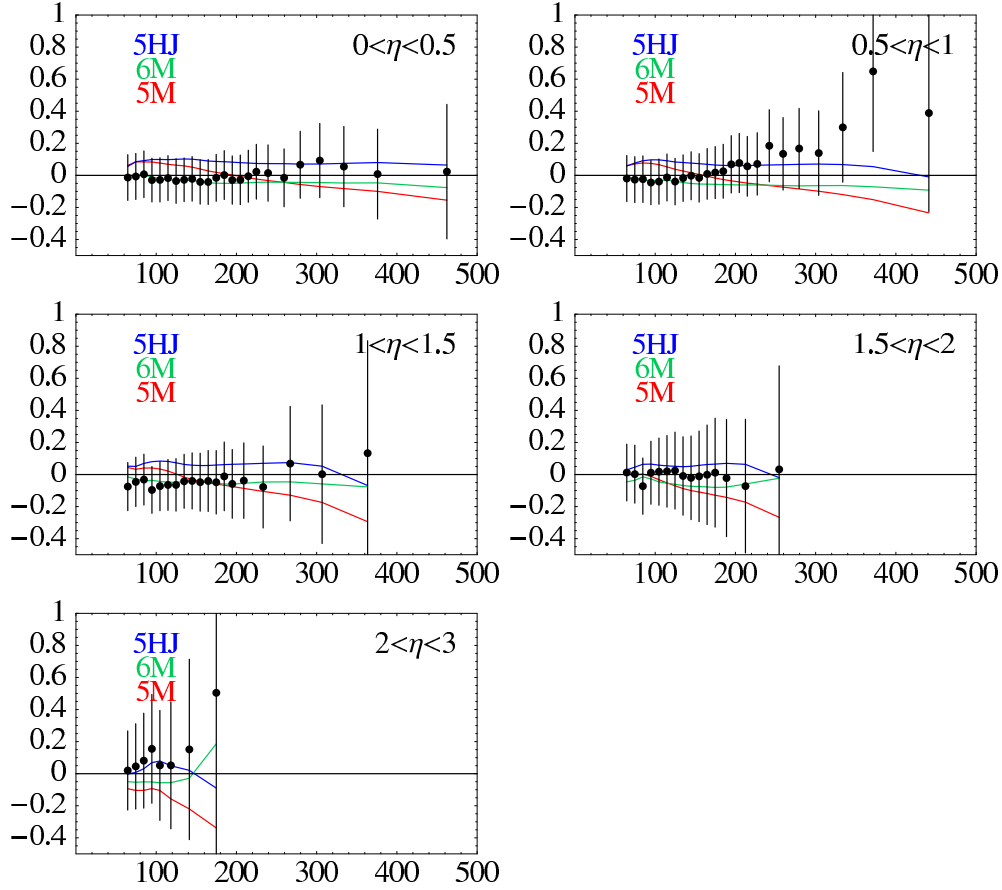}
\end{center}
\caption{The D\O\ inclusive jet cross section versus $E_{T}$
for five rapidity bins compared to NLO predictions
using the CTEQ6.1M PDF's.
The abscissa is $E_{T}$ in GeV.
The ordinate is (data $-$ theory)/theory.
The curves show the CTEQ5M, CTEQ5HJ, and CTEQ6M
predictions,
as fractional differences compared to CTEQ6.1M.
The error bars are the statistical and systematic errors
combined in quadrature.
\label{fig:D0vsCTEQ6}}
\end{figure}

The comparison between the D\O\ data
for the 5 different rapidity intervals,
and the predictions using the CTEQ6.1M PDF's,
is shown in Fig.\,\ref{fig:D0vsCTEQ6}.
The theory predictions for CTEQ6M, CTEQ5M and
CTEQ5HJ parton distributions are also shown.
The greater statistical power that results from including
the new D\O\ jet data, and the fact that all of the rapidity
intervals prefer a larger gluon at high $x$, results in the
increases in the cross sections found using NLO
predictions based on CTEQ6M or CTEQ6.1M rather than CTEQ5M.

This point deserves further emphasis.
It is crucial  to measure the jet cross
section over a wide rapidity range: any new physics should
contribute mostly to the central rapidity region, while a PDF explanation
will contribute to all rapidity ranges. This will be addressed 
in more detail below.

\begin{figure}[ht]
\begin{center}
\parbox{0.49\textwidth}
{\includegraphics[width=0.49\textwidth]{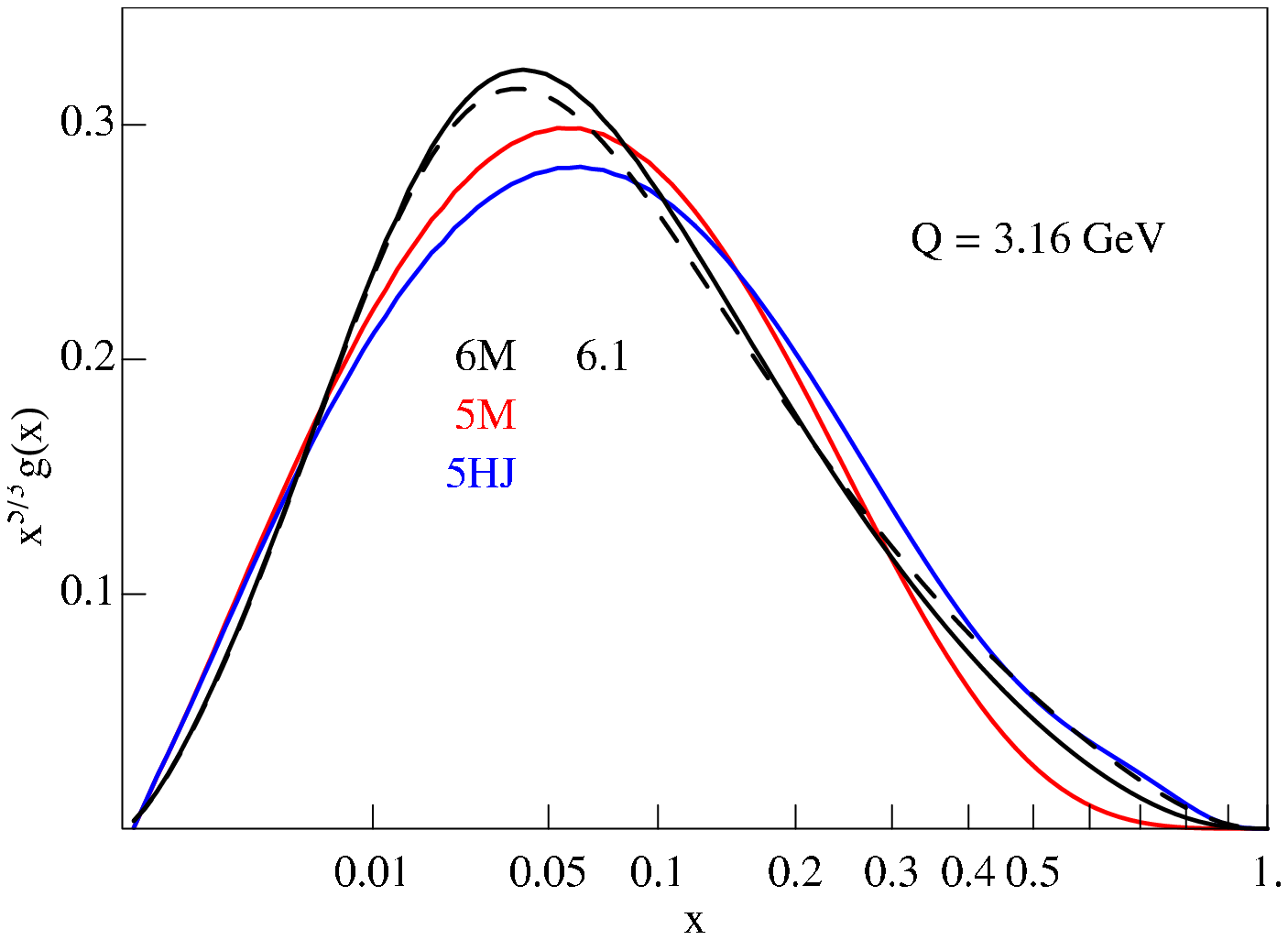}}
\hfill
\parbox{0.49\textwidth}
{\includegraphics[width=0.49\textwidth]{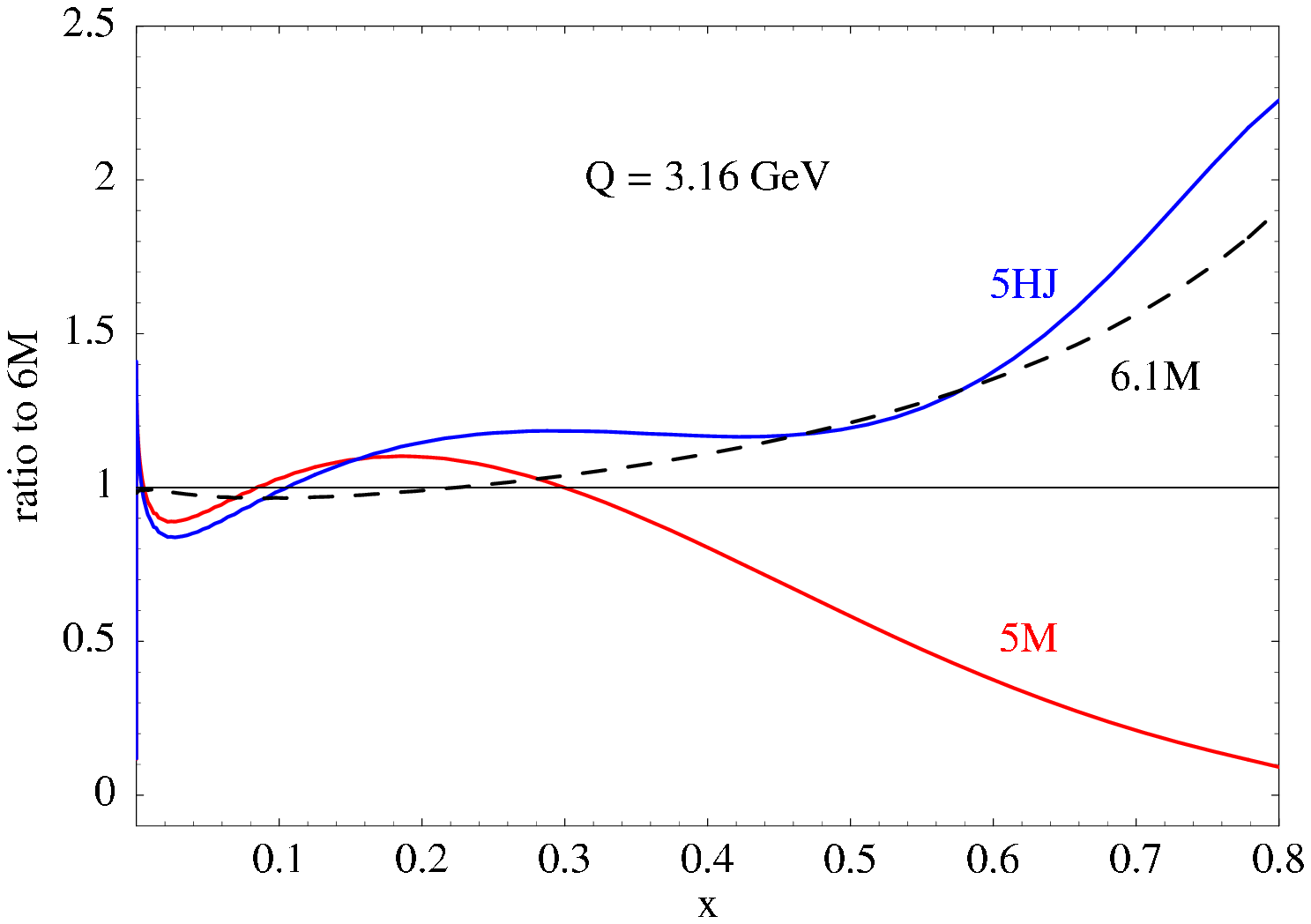}}
\end{center}
\caption{Left: The CTEQ5M, CTEQ5HJ, and CTEQ6M gluon distributions
at $Q^{2}=10$\,GeV$^{2}$.
Right: The ratios of CTEQ5M and CTEQ5HJ gluon distribution
to that of CTEQ6M.
The dashed curves show the CTEQ6.1M gluon distribution.
\label{fig:comparePDFs}}
\end{figure}

As discussed in Ref.~\cite{cteq6}, it is important to have a
parameterization for the PDF's that is flexible enough to allow an 
increase in the high $x$ gluon,
without modifying too much the shape of the gluon distribution
at lower values of $x$.
A comparison of the CTEQ5M, CTEQ5HJ and CTEQ6M gluon distributions is
shown in Fig.\,\ref{fig:comparePDFs}.
As can be observed, the CTEQ6M gluon at high $x$ lies between the
gluons in CTEQ5M and in CTEQ5HJ, just as the CTEQ6M predictions for
the inclusive jet cross sections at high $E_{T}$ lie between the
predictions using the two other PDF's. 
The CTEQ6.1M gluon distribution is shown
by dashed curves in Fig.\,\ref{fig:comparePDFs}.
At the highest $x$ values the function resembles CTEQ5HJ,
and in the intermediate $x$ range it resembles CTEQ6M.

\begin{figure}[ht]
\begin{center}
\includegraphics[width=0.7\textwidth,angle=270]{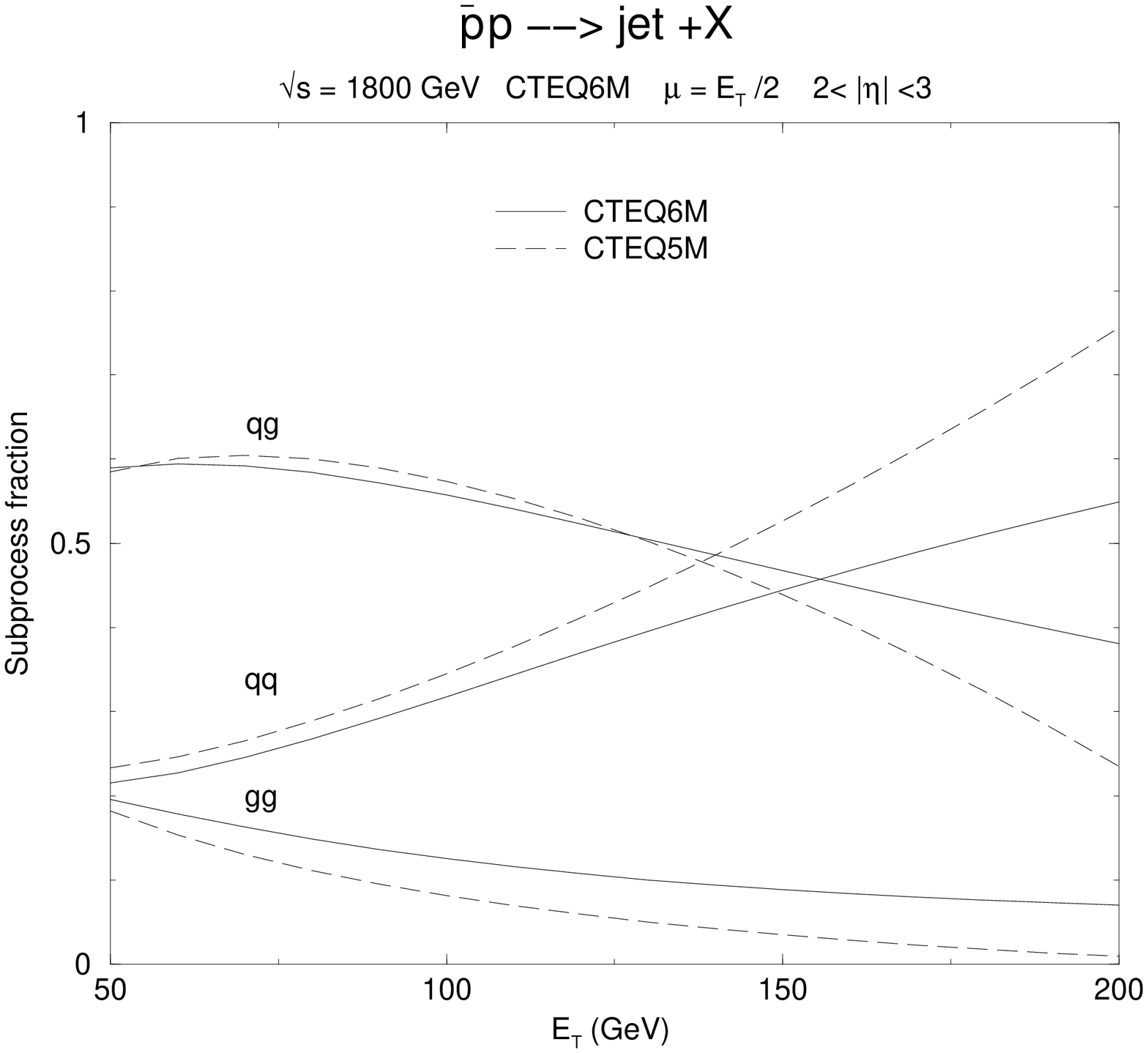}
\end{center}
\caption{The subprocess contributions to jet production
in the forward rapidity region.
\label{fig:d0five5m6m}}
\end{figure}

The subprocess contributions to jet production are shown
in the next figures,
for the central rapidity bin (Fig.\,\ref{fig:d0one5m6m})
and the forward rapidity bin (Fig.\,\ref{fig:d0five5m6m}),
for the parton distributions CTEQ5M and CTEQ6M.

The dominant difference between the CTEQ5M and CTEQ6M sets is 
the increased gluon distribution in CTEQ6M at large values 
of $x$; the quark distributions are nearly unchanged. This shows up 
as a decrease in the fractional contributions of the quark-quark 
subprocesses and as increases in the fractional contributions of the 
quark-gluon and gluon-gluon subprocesses.
It is interesting to note that the $gq$ scattering subprocess
is even more important at high $E_{T}$ for the highest rapidity
bin than for the central rapidity bin. 

\begin{figure}[ht]
\begin{center}
\includegraphics[width=0.7\textwidth,angle=270]{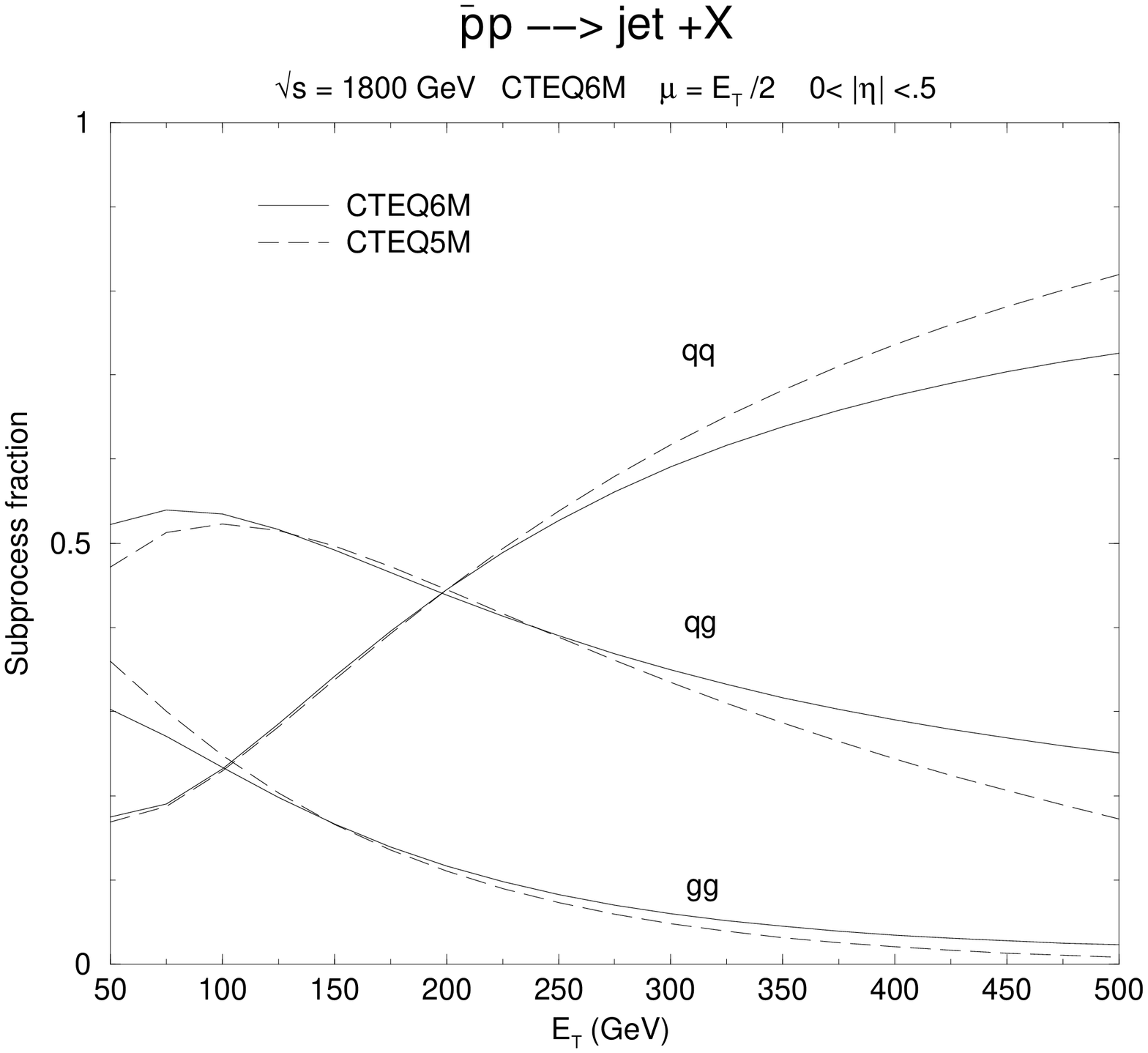}
\end{center}
\caption{The subprocess contributions to central jet production.
\label{fig:d0one5m6m}}
\end{figure}

The cross section for production of high-$E_{T}$ jets
at the Tevatron depends on the parton distributions
at large $x$.
The relevant range of $x$ may be estimated by considering
the leading-order $2\rightarrow{2}$ kinematics.
Let $x_{1}$ and $x_{2}$ be the momentum fractions
of the two incoming partons.
The outgoing parton 4-momenta are
\begin{eqnarray}
p_{1}^{\prime\mu} &=& \left(p_{T}\cosh y_{1}, \vec{p}_{T},
p_{T}\sinh y_{1} \right),
\\
p_{2}^{\prime\mu} &=& \left(p_{T}\cosh y_{2}, -\vec{p}_{T},
p_{T}\sinh y_{2} \right);
\end{eqnarray}
the outgoing partons have rapidities $y_{1}$ and $y_{2}$,
and transverse momenta $\vec{p}_{T}$ and $-\vec{p}_{T}$,
respectively.
If the first parton is identified with the observed jet
then the jet rapidity is $y_{j}=y_{1}$ and its transverse
momentum is $\vec{p}_{T}$.
The $2\rightarrow{2}$ kinematics determines the relation
between the momentum fractions and outgoing parton variables,
\begin{eqnarray}
x_{1} &=& \frac{p_{T}}{\sqrt{s}}\left( e^{y_{j}}+e^{y_{2}}\right),
\\
x_{2} &=& \frac{p_{T}}{\sqrt{s}}\left( e^{-y_{j}}+e^{-y_{2}}\right).
\end{eqnarray}
These parametric equations yield a curve in the ($x_{1},x_{2}$)
plane, parametrized by the rapidity $y_{2}$ of the second parton,
for the specified jet variables.

\begin{figure}[ht]
\begin{center}
\parbox{0.49\textwidth}
{\includegraphics[width=0.49\textwidth]{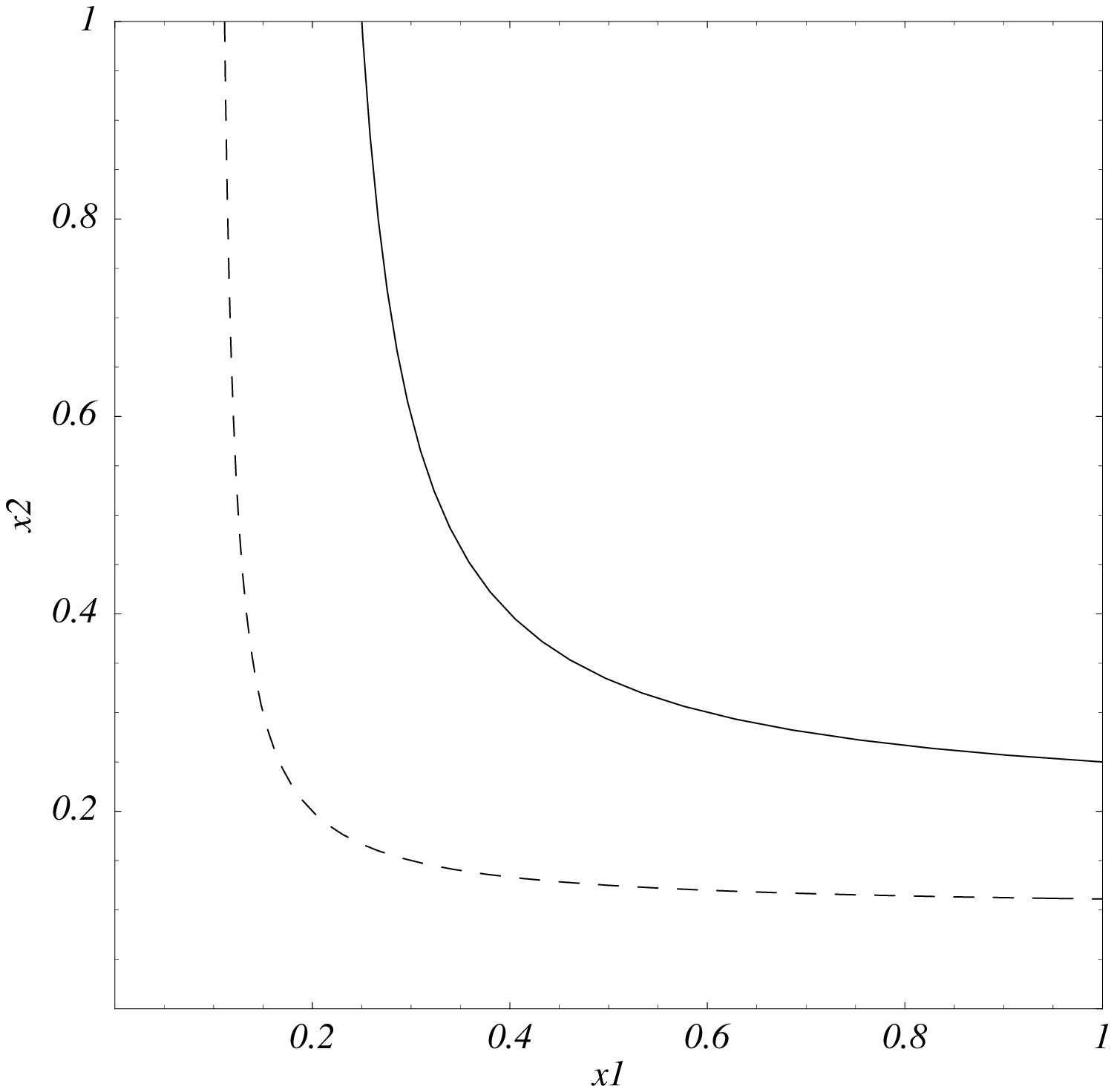}}
\hfill
\parbox{0.49\textwidth}
{\includegraphics[width=0.49\textwidth]{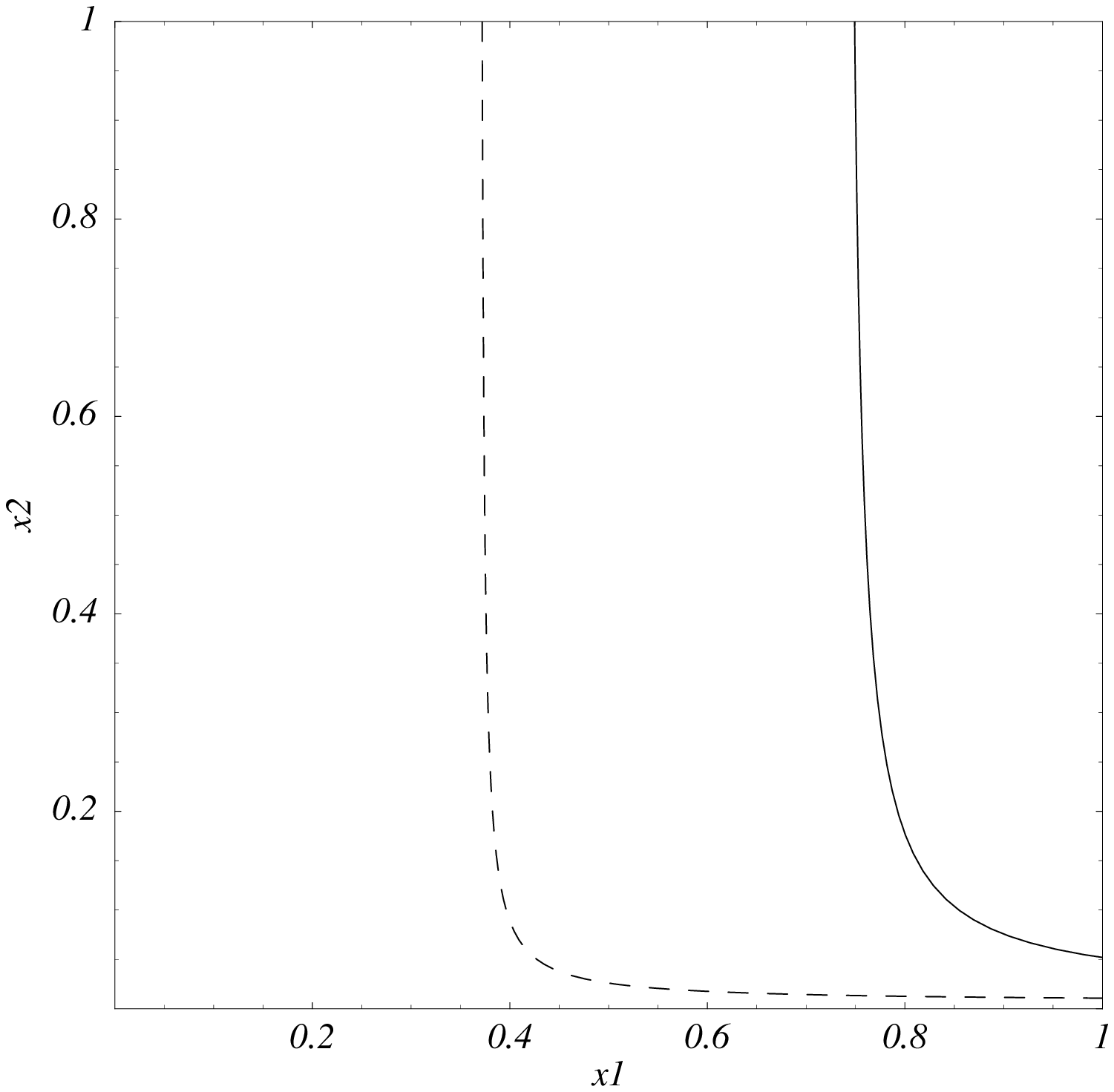}}
\end{center}
\caption{Left: Parton momentum fractions $x_{1}$ and $x_{2}$
for $y_{j}=0$;
the solid and dashed curves are for $E_{T}=400$ and $200$\,GeV,
respectively.
Right: Parton momentum fractions $x_{1}$ and $x_{2}$
for $y_{j}=2$;
the solid and dashed curves are for $E_{T}=200$ and $100$\,GeV,
respectively.
\label{fig:x1x2hyperbolas}}
\end{figure}

Figure \ref{fig:x1x2hyperbolas} shows the $(x_{1},x_{2})$
values for central rapidity ($y_{j}=0$) on the left and
forward rapidity ($y_{j}=2$) on the right.
In the central case, the solid curve is the
locus of $(x_{1}, x_{2})$ points
for $p_{T}=400$\,GeV and the dashed curve
is for $p_{T}=200$\,GeV.
Both $x_{1}$ and $x_{2}$ must be $\gtrsim 0.25$
to produce the high-$p_{T}$ jet.
In the forward case the solid curve corresponds to
$p_{T}=200$\,GeV and the dashed curve to $p_{T}=100$\,GeV.
Here one parton must have large momentum fraction
to produce the jet at large $p_{T}$ and forward rapidity.
The second parton can have a low value of $x$,
leading to the second jet being close in rapidity to the first jet,
or a larger value of $x$, leading to configurations where the two
jets are on opposite sides of the detector.
Consider the contribution of the quark-gluon scattering subprocesses. 
For the configuration where one $x$ is small and the other is large, 
the smaller $x$ value most likely corresponds to a gluon and the larger 
to a quark.
In this case the difference between the CTEQ5M and CTEQ6M distributions 
is small. On the other hand, for the configuration where both $x$ 
values are large the increased gluon in CTEQ6M will enhance the 
cross section. The conclusion is that the increase in the cross section 
when going from CTEQ5M to CTEQ6M occurs in configurations where the 
forward jet is balanced in $p_T$ by a jet with roughly the opposite rapidity. 


Next, consider the question of where in the jet phase space one is 
most likely to observe signs of compositeness or other new physics. 
The angular distributions of the dominant QCD subprocesses are sharply 
peaked at small scattering angles, whereas the models usually used to 
estimate compositeness have much flatter distributions. This means 
that the signal/QCD ratio will be largest for 90$^{\circ}$ scattering 
in the parton-parton rest frame. In the overall hadron-hadron frame, this 
corresponds both jets having the same rapidity. Using the lowest order 
kinematics described previously, the squared dijet mass is given by 
$$M_{jj}^2 = 2 p_T^2[1+\cosh(y_1-y_2)].$$

\noindent Setting $y_1 = y_2$ yields $M_{jj}^2=4p_T^2.$ This shows that 
maximizing the dijet mass means maximizing $p_T$. Since the $p_T$ reach 
is maximized in the central region, this is the region where the highest 
dijet masses can be reached with the lowest background for new physics 
signals (at least those with a flat angular distribution). The 
model discussed for compositeness later in this paper gives an explicit 
example of this behavior. On the other hand, as shown by the kinematic 
examples in Figs.\, \ref{fig:x1x2hyperbolas}, modifications of the gluon 
distribution at high values of $x$ will affect the high-$p_T$ jet cross 
section at all rapidities.

\clearpage
\setcounter{section}{1}

\section{Uncertainty analysis of the cross section for
inclusive jet production at the Tevatron}

As discussed in the previous section,
the cross section for inclusive jet production has been
measured by the CDF and D\O \ collaborations
at the Tevatron.
These results have been used in the latest CTEQ global
analysis of PDF's, leading to the CTEQ6 (and now 6.1M)
parton distributions.
In this section we analyze the uncertainty of the theoretical
QCD cross section due to PDF uncertainty.

\subsection{The Hessian method of uncertainty analysis}

The Hessian method of uncertainty analysis has been described in detail
previously~\cite{PSTHessian}.
For completeness, we summarize the method briefly.
The parton distributions are constructed by a method of
chi-square minimization with fitting of systematic errors.
A chi-square function $\chi^{\prime \, {2}}$ is defined by
\begin{eqnarray}
\chi^{\prime \, {2}} &=& \sum_{e} \chi_{e}^{2}(a,r)
\\
\chi_{e}^{2}(a,r) &=& \sum_{i}
\frac{\left[D_{i}-\sum_{k}r_{k}\beta_{ki} - T_{i}(a)\right]^{2}}
{\alpha_{i}^{2}}
+\sum_{k}r_{k}^{2}
\end{eqnarray}
where $e$ labels an experimental data set
and $i$ labels a data point in the data set.
$D_{i}$ is the data value, $\alpha_{i}$ is the uncorrelated error,
and $\beta_{ki}$ is the $k$th correlated systematic error;
these numbers are published by the experimental collaboration.
Then $T_{i}(a)$ is the theoretical value, a function of a set
of $n$ PDF parameters, $\{a_{1}, \dots , a_{n}\}$.
Also, $\{r_{k}\}$ is a set of Gaussian random variables
and $r_{k}\beta_{ki}$ is a (correlated) shift
applied to $D_{i}$ to represent the $k$th systematic error.
We minimize the function $\chi^{\prime \, {2}}(a,r)$ with
respect to both the PDF parameters $\{a\}$ and the
systematic shift variables $\{r_{k}\}$.
The result yields both the standard PDF model
with parameters $\{a_{0}\}$, and the optimal shifts
$\{\widehat{r}_{k}\}$ to bring theory and data
into agreement.
This minimum of $\chi^{\prime \, {2}}$
represents the best fit to the data~\cite{cteq6}.

The Hessian matrix is
\begin{equation}
H_{ij}=\frac{1}{2}\frac{\partial^{2}\widehat{\chi}^{2}}
{\partial{a}_{i}\partial{a}_{j}}
\qquad \mbox{where} \qquad
\widehat{\chi}^{2}(a) \equiv \chi^{\prime \, {2}}(a,\widehat{r}(a)).
\end{equation}
This matrix determines the behavior of $\widehat{\chi}^{2}(a)$
in the neighborhood of the minimum.
We have developed an iterative method
for calculating the Hessian accurately,
leading to an improvement in the
$\chi^{2}$ minimization~\cite{PSTIterative}.
The point $\{a_{0}\}$ in the $n$-dimensional parameter
space---where $\widehat{\chi}^{2}(a)$ is minimum---is
the best fit to the global data set.
However, points in some small neighborhood of $\{a_{0}\}$
are also acceptable fits.
The variation within the acceptable neighborhood,
of a PDF or a physical prediction,
is our measure of the uncertainty of the quantity.

In the Hessian method of uncertainty analysis,
we first define $n$ special directions in the parameter space,
namely the $n$ eigenvectors of the Hessian.
Then for each eigenvector we have two displacements
from $\{a_{0}\}$
(in the $+$ and $-$ directions along the vector)
denoted $\{a^{+}_{i}\}$ and $\{a^{-}_{i}\}$
for the $i$th eigenvector.
At these points,
$\widehat{\chi}^{2}=\widehat{\chi}_{0}^{2}+T^{2}$
where $\widehat{\chi}_{0}^{2}=\widehat{\chi}^{2}(a_{0})$
$=$ the minimum,
and $T$ is a parameter called {\em the tolerance}.
We consider any PDF set with
$\widehat{\chi}^{2}-\widehat{\chi}_{0}^{2}<T^{2}$
to be an acceptable fit to the global data set.
In particular, the $2n$ PDF sets $\{a_{i}^{\pm}\}$,
which are called the eigenvector basis sets, span the
parameter space in the neighborhood of the minimum.

The appropriate choice of tolerance $T$ cannot be decided
without a further, more detailed, analysis of the quality
of the global fits.
After studying a number of examples~\cite{PSTHessian, PSTLagrange}
we conclude that a rather large tolerance, $T \sim 10$,
represents a realistic estimate of the PDF uncertainty.\footnote{%
Different choices of the tolerance have been made by other groups,
e.g., $T\sim \sqrt{40}$ by MRST~\cite{mrst2002}.}

Any quantity $X$ that depends on PDF's
has a predicted value $X_{0} = X(a_{0})$ and a range of
uncertainty $\delta{X}$.
A simple measure of $\delta{X}$ is the spread of the values
of $X(a^{\pm}_{i})$ for the $2n$ eigenvector basis sets.
However, a more complete uncertainty range is between the
minimum and maximum values of $X$ for all points
with $\widehat{\chi}^{2}-\widehat{\chi}_{0}^{2}<T^{2}$.
It can be shown that in a linear approximation
these bounds are $X_{0}\pm \delta{X}$ where
\begin{equation}
\left(\delta{X}\right)^{2}
= T^{2} \sum_{i,j}
\left(H^{-1}\right)_{ij}
\frac{\partial{X}}{\partial{a}_{i}}
\frac{\partial{X}}{\partial{a}_{j}};
\end{equation}
or, in terms of the eigenvector basis sets,
\begin{equation}\label{eq:ME1}
\left(\delta{X}\right)^{2}
= \frac{1}{4}\sum_{k=1}^{n}
\left[X(a^{+}_{i})-X(a^{-}_{i})\right]^{2}.
\end{equation}
We refer to (\ref{eq:ME1}) as the
Master Equation for the uncertainty of $X$
in the Hessian method~\cite{PSTHessian}.
Equation (\ref{eq:ME1}) is based on a linear approximation:
$\widehat{\chi}^{2}(a)$ is assumed
to be a quadratic function of the parameters $\{a\}$,
and $X(a)$ is assumed to be linear.
We find that these approximations are not strictly valid,
so instead we calculate asymmetric bounds by
\begin{equation}\label{eq:MEasy}
\delta{X}_{\pm} = \sum_{k=1}^{n}
\left[X(a^{\pm}_{i})-X(a_{0})\right]^{2};
\end{equation}
the range of uncertainty of $X$ is
$(X_{0}-\delta{X}_{-},X_{0}+\delta{X}_{+})$.
(In (\ref{eq:MEasy})
$\{a_{i}^{+}\}$ and $\{a_{i}^{-}\}$ are the
displaced points where
$X>X_{0}$ and $X<X_{0}$, respectively.)

\begin{figure}[ht]
\begin{center}
\parbox{0.49\textwidth}
{\includegraphics[width=0.49\textwidth]{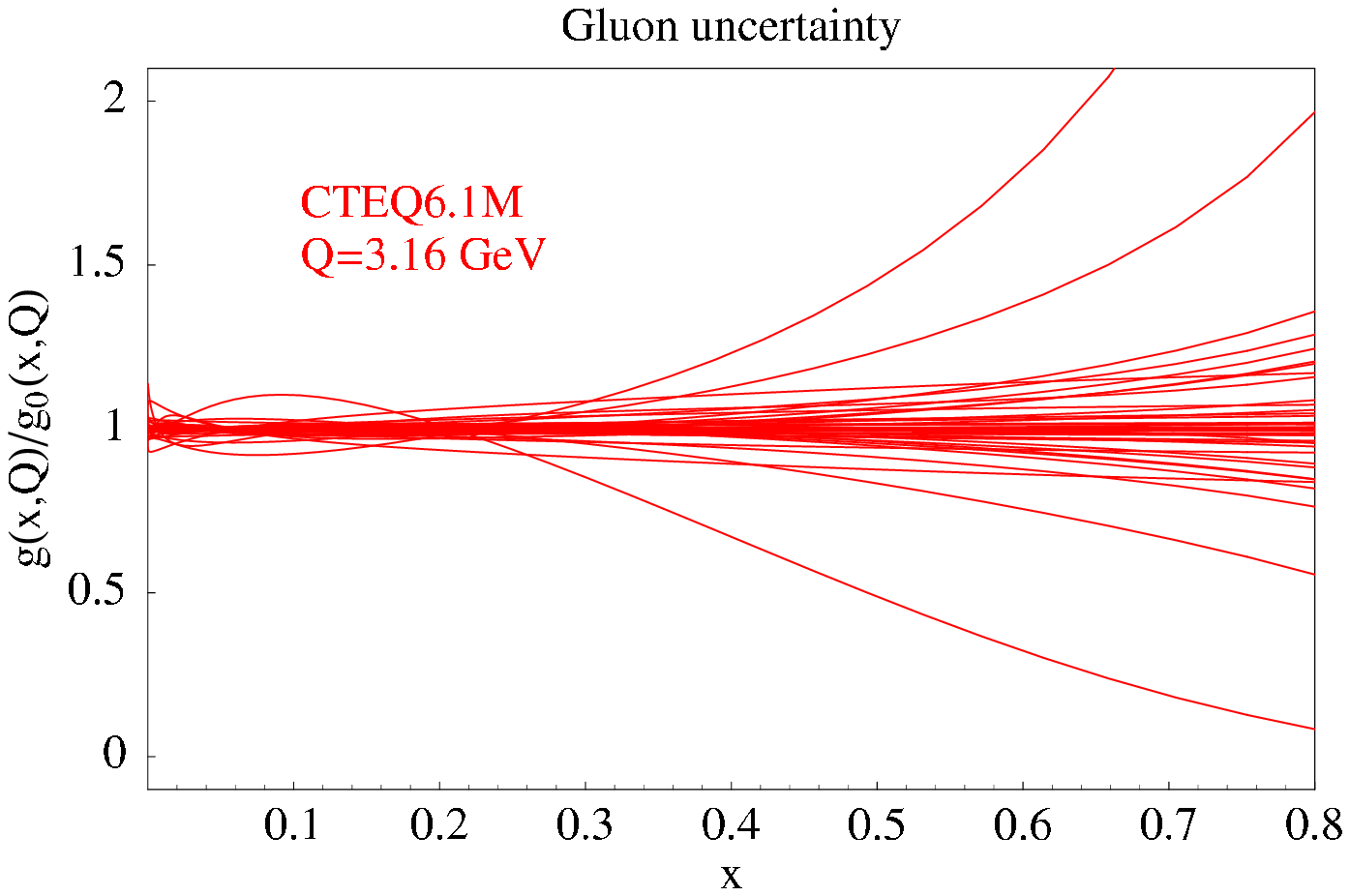}}
\hfill
\parbox{0.49\textwidth}
{\includegraphics[width=0.49\textwidth]{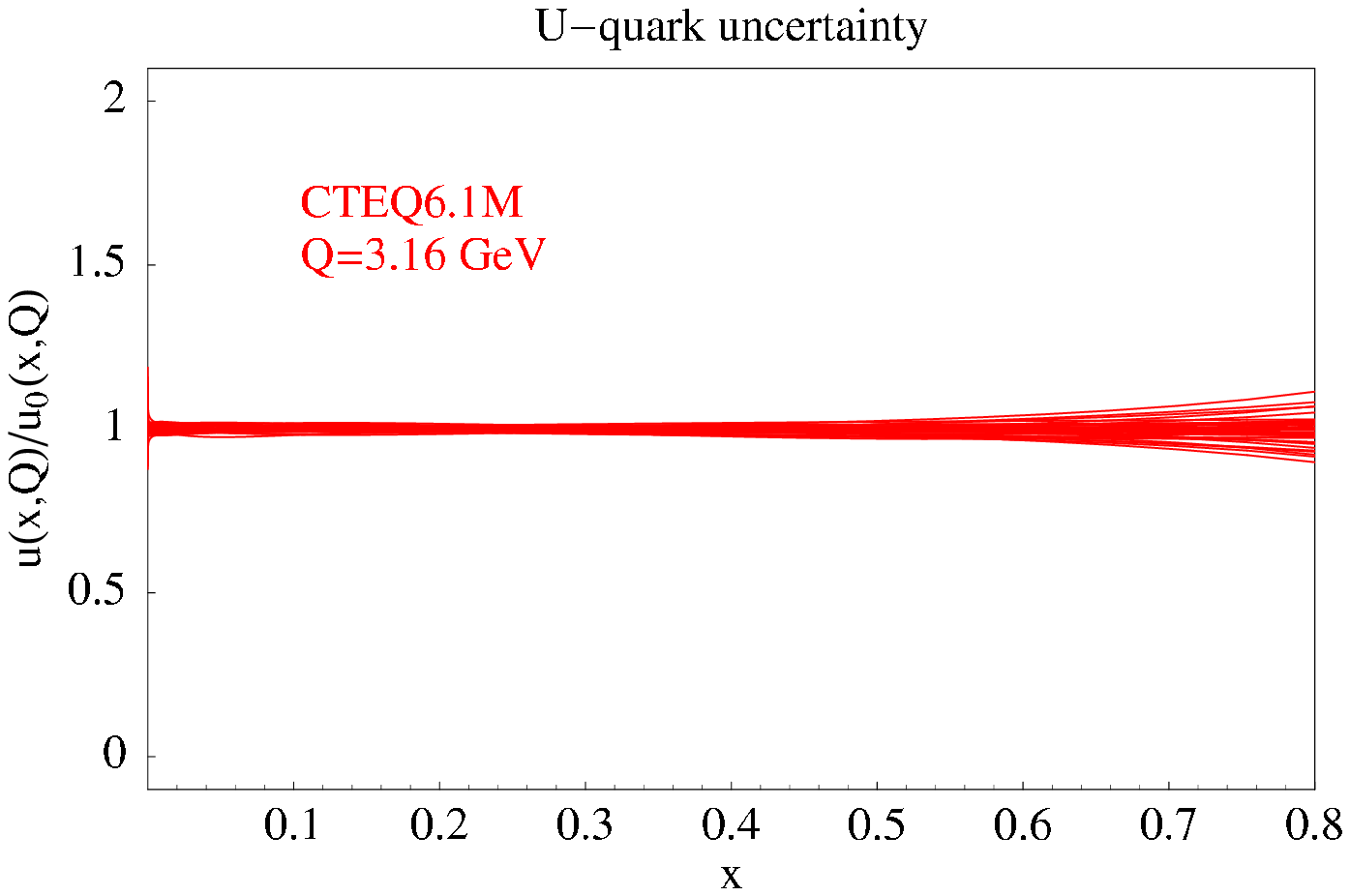}}
\end{center}
\caption{PDF uncertainties.
The curves are the ratios of the 40 eigenvector basis
sets to the standard set (CTEQ6.1M).
The largest variations of the gluon function are
the $+$ and $-$ displacements along eigenvector 15.
\label{fig:PDFunc}}
\end{figure}

\begin{figure}[ht]
\begin{center}
\includegraphics[width=0.49\textwidth]{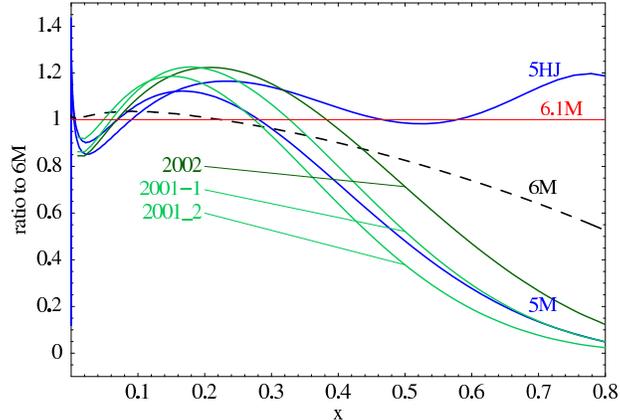}
\end{center}
\caption{Comparison of several standard PDF sets.
\label{fig:StdSets}}
\end{figure}

The uncertainties on the gluon and $u$-quark
distributions are shown in Fig.\,\ref{fig:PDFunc},
for $Q^{2}=10$\,GeV$^{2}$.
The $u$-quark distribution is tightly constrained for
$x \lesssim 0.8$, whereas the uncertainty on the gluon
distribution grows to a factor of 2 or larger for $x$
values greater than $\sim 0.4$. 
For comparison, the CTEQ5M, CTEQ5HJ, MRST2001~\cite{mrst2001} and 
MRST2002~\cite{mrst2002} PDF's are shown, compared to CTEQ6.1M,
in Fig.\,\ref{fig:StdSets}, for $Q^{2}=10$\,GeV$^{2}$.

\subsection{The Tevatron jet cross section}

\begin{figure}[ht]
\begin{center}
\parbox{0.49\textwidth}
{\includegraphics[width=0.49\textwidth]{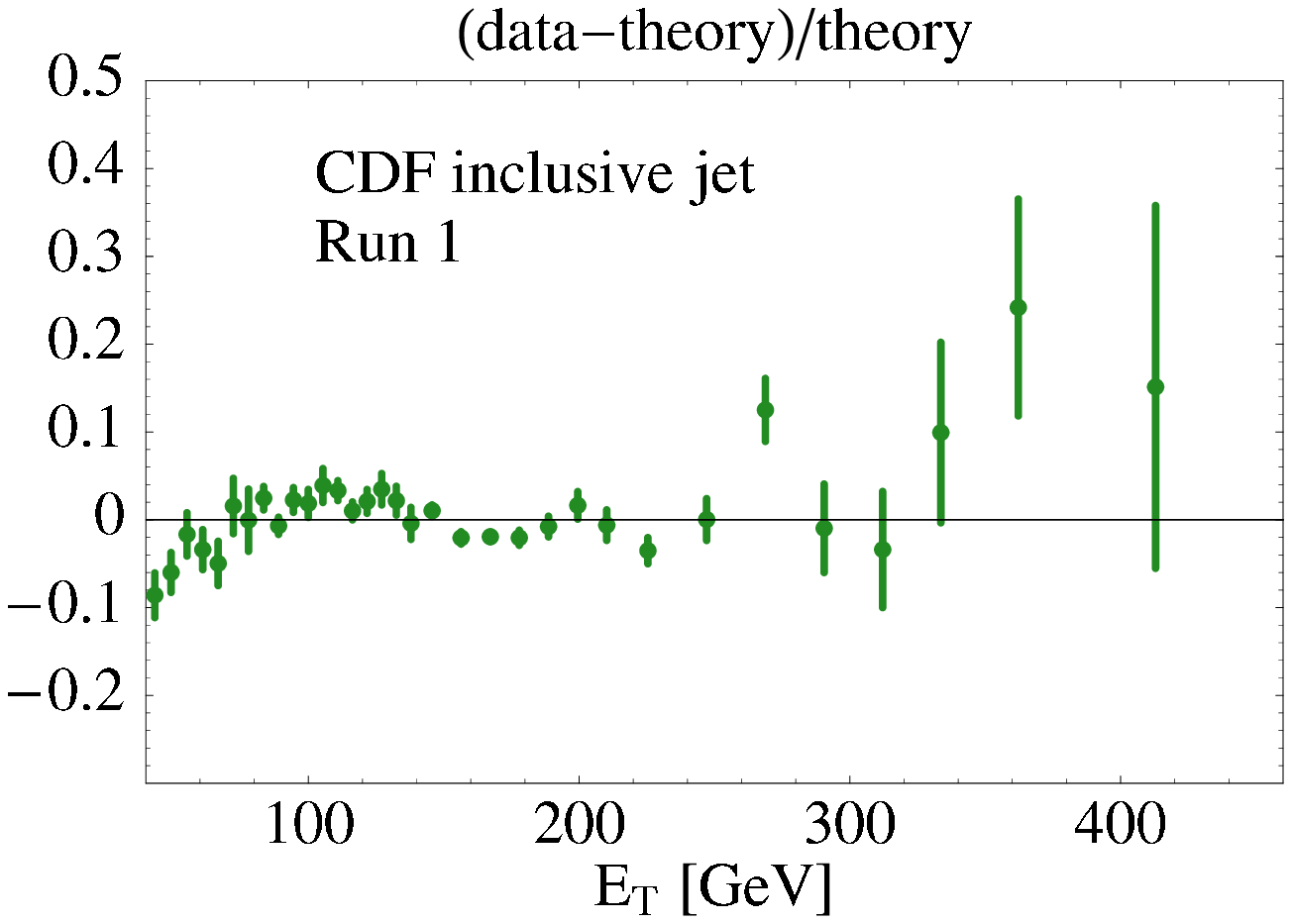}}
\parbox{0.49\textwidth}
{\includegraphics[width=0.49\textwidth]{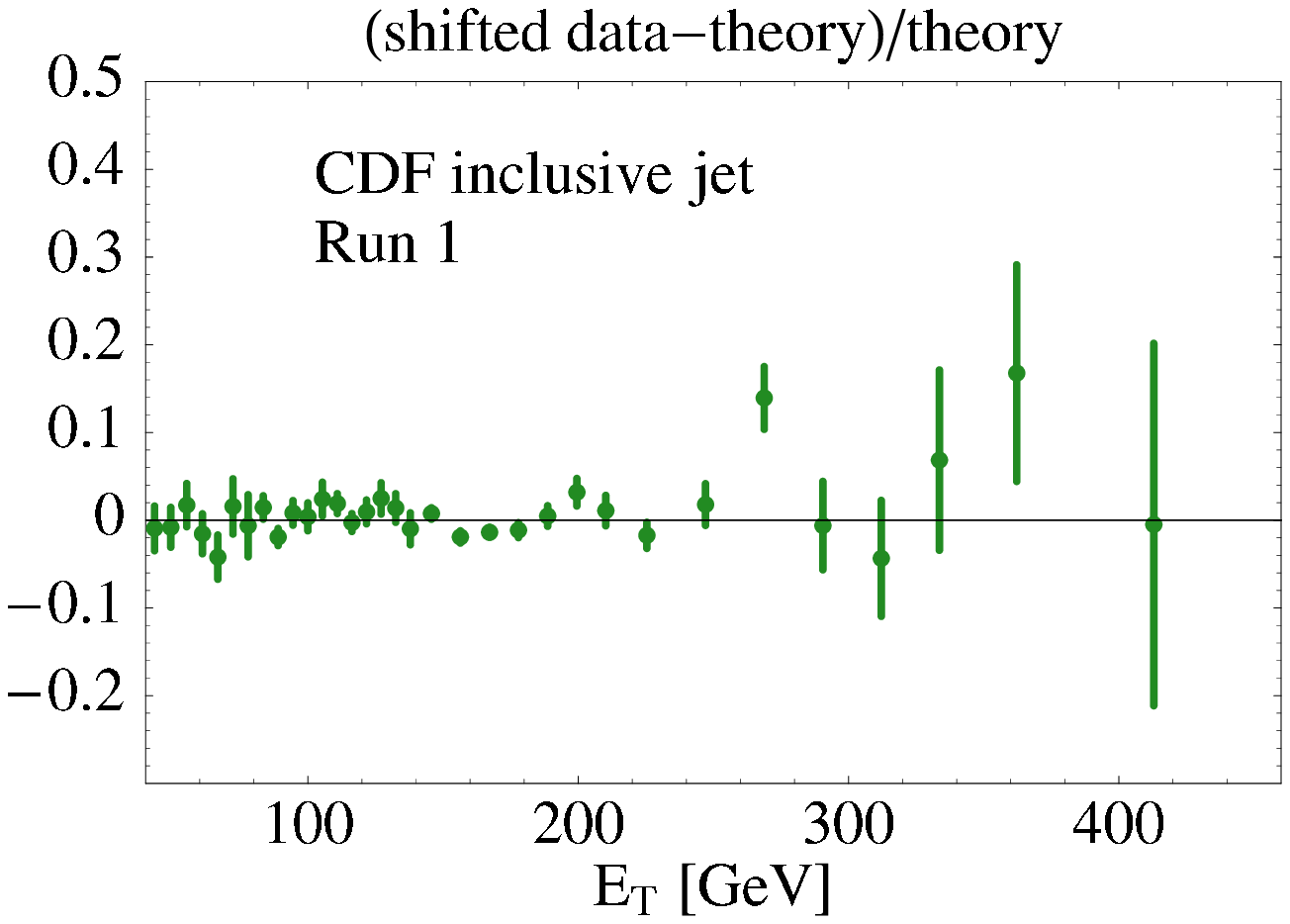}}
\end{center}
\caption{The left panel shows $(D_{i}-T_{i})/T_{i}$,
the fractional difference between data and theory.
The right panel shows $(\overline{D}_{i}-T_{i})/T_{i}$
where $\overline{D}_{i}$ is the data with systematic
shifts subtracted.
The error bars are the statistical errors only.
\label{fig:DRSjetfig0}}
\end{figure}

To illustrate our method, and the importance of the systematic
errors, Fig.\,\ref{fig:DRSjetfig0} shows the fractional difference
between the data obtained by CDF in Run 1b, and the theory
with CTEQ6.1M partons.
The error bars are the statistical errors only.
The left panel is $(D_{i}-T_{i})/T_{i}$.
Note that there are some systematic differences between the CDF jet
data and the theory, both in terms of normalization
and in shape.\footnote{%
It is important to note all Run 1 CDF cross sections are intrinsically
larger than comparable D\O\ cross sections by approximately 6\% due
to different assumptions regarding the 
total inelastic cross section at 1.8 TeV~\cite{cdf_d0_sigma}.}
The right panel shows $(\overline{D}_{i}-T_{i})/T_{i}$
where $\overline{D}_{i}$ is the shifted data,
\begin{equation}
\overline{D}_{i} = D_{i}-\sum_{k=1}^{n} \widehat{r}_{k}\beta_{ki};
\end{equation}
i.e., the optimal systematic shifts determined by the global fit
have been subtracted from the data values.
There is no systematic difference between $\overline{D}_{i}$
and $T_{i}$, as expected.
The sizes of the systematic shifts are, as expected, comparable
to the standard deviations published by the CDF collaboration;
that is, $\widehat{r}_{k}$ is of order $1$ for each of the
systematic errors.

\begin{figure}[ht]
\begin{center}
\includegraphics[width=0.75\textwidth]{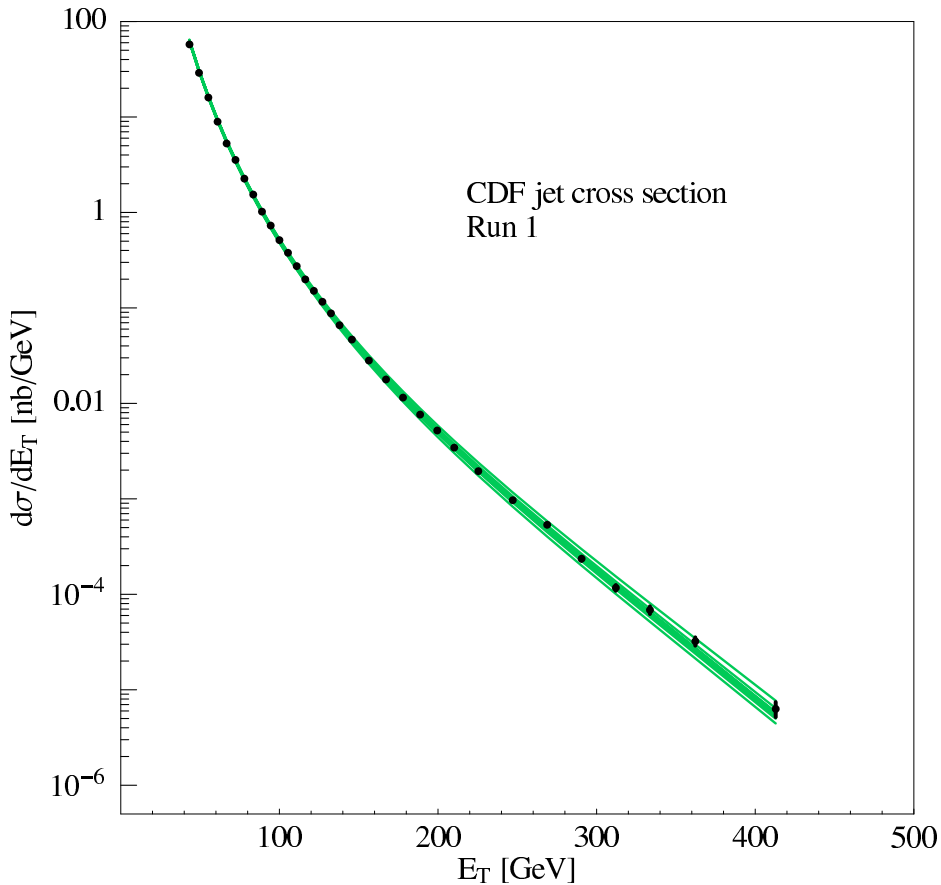}
\end{center}
\caption{Calculations of $d\sigma/dE_{T}$
for the 40 eigenvector sets.
\label{fig:CDFjetA}}
\end{figure}

Figure \ref{fig:CDFjetA} shows the cross section
for inclusive jet production at the Tevatron,
integrated over the rapidity range $0.1<|\eta|<0.7$,
as a function of jet transverse energy $E_{T}$,
for CTEQ6.1M partons and for the 40 eigenvector sets.
(The rapidity interval is that for which the cross
section has been published by the CDF collaboration.)
The spread of the 41 curves in Fig.\,\ref{fig:CDFjetA}
is a simple estimate of the PDF uncertainty.
The CDF data points are superimposed on the calculation.

\begin{figure}[ht]
\begin{center}
\includegraphics[width=0.75\textwidth]{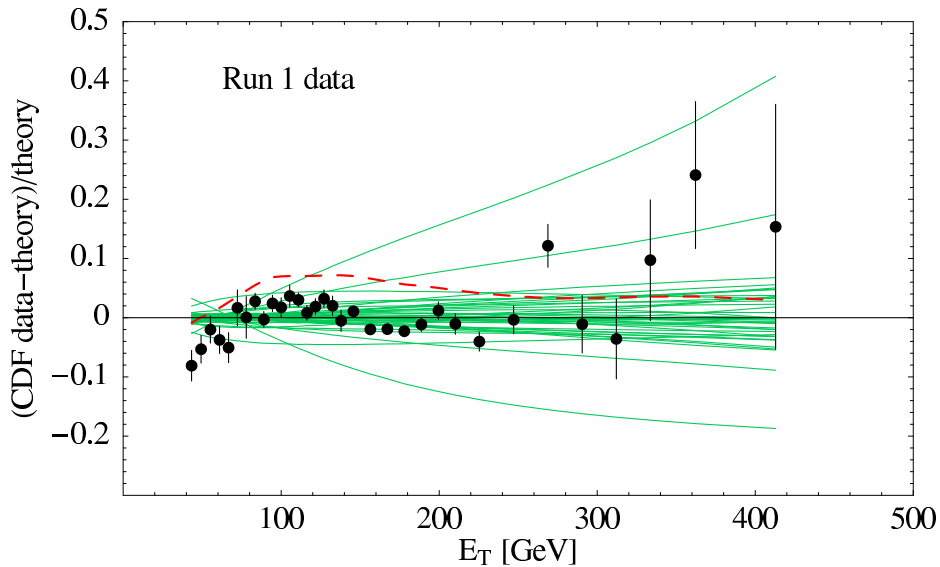}
\end{center}
\caption{Calculations for the 40 eigenvector basis sets,
plotted as fractional differences compared to CTEQ6.1M.
The points are the CDF measurements.
The error bars are statistical errors only;
the systematic shifts are not subtracted from the data.
The dashed curve is CTEQ5HJ compared to CTEQ6.1M.
\label{fig:CDFjetB}}
\end{figure}

Figure \ref{fig:CDFjetB} shows the fractional differences
$\left[T(a_{i}^{\pm})-T(a_{0})\right]/T(a_{0})$
between the 40 sets and the standard prediction (CTEQ6.1M).
The points are the CDF data compared to CTEQ6.1M;
the error bars are statistical
errors and the data points are plotted
{\em without} the systematic shifts.
Also shown is the prediction using CTEQ5HJ partons.
The range of uncertainty encompasses both the CDF data
and the CTEQ5HJ 
predictions.

\begin{figure}[ht]
\begin{center}
\includegraphics[width=0.75\textwidth]{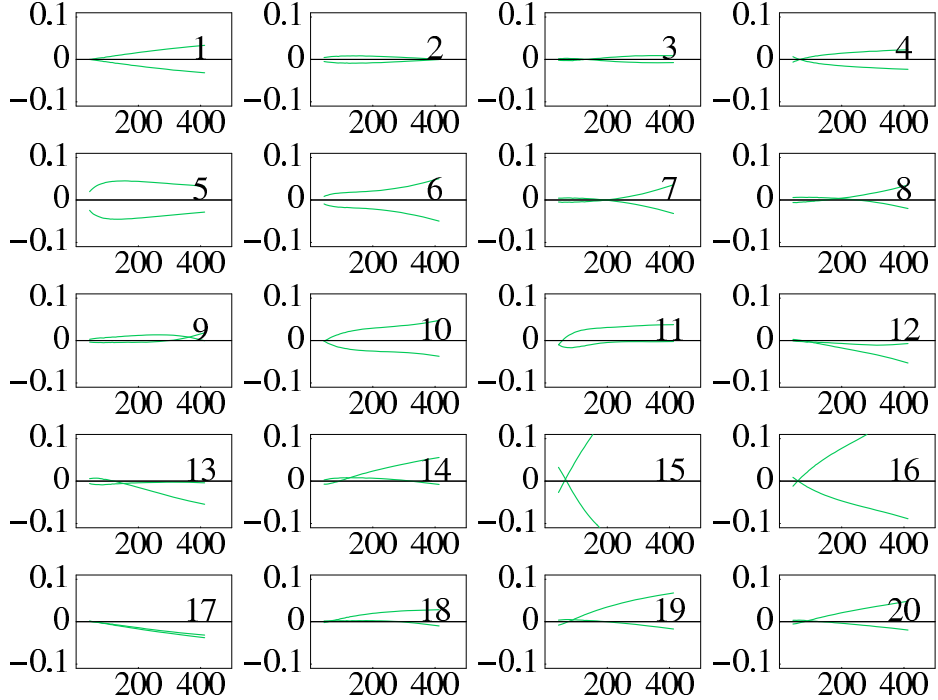}
\end{center}
\caption{The cross sections for $+$ and $-$ displacements
along the 20 eigenvectors, plotted as fractional differences
compared to CTEQ6.1M.
\label{fig:CDFjetC}}
\end{figure}

Figure \ref{fig:CDFjetC} shows the 40 eigenvector
basis sets separately.
For each eigenvector the two curves are
the positive and negative displacements.
The Hessian method assumes that the variations
are approximately linear.
If the linear approximation were strictly valid
the curves would be mirror images about zero.
Figure \ref{fig:CDFjetC} implies that the linear
approximation is reasonably accurate for  most
eigenvectors, especially those corresponding to the
ten largest eigenvalues (1 -- 10).

\begin{figure}[ht]
\begin{center}
\includegraphics[width=0.75\textwidth]{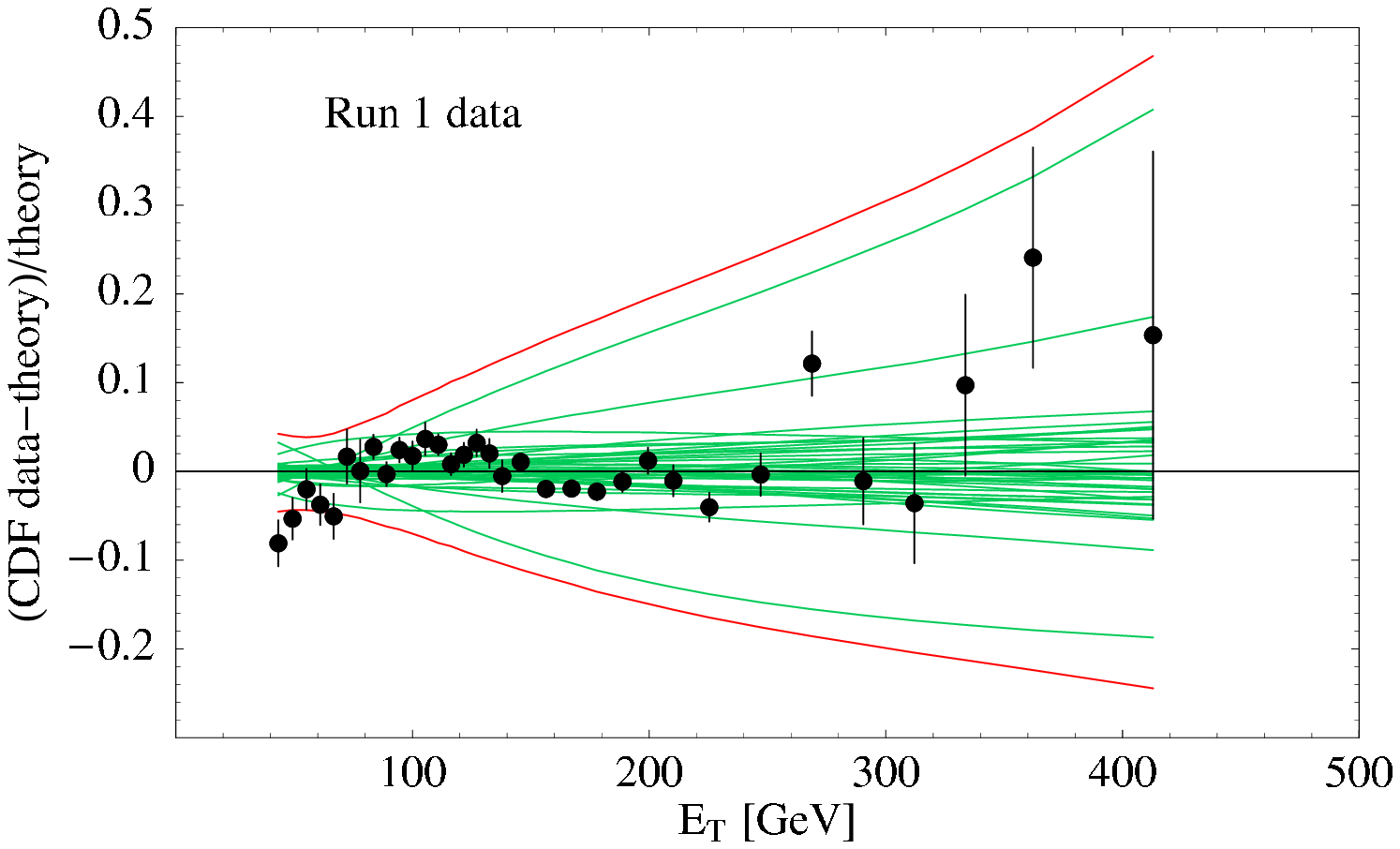}
\end{center}
\caption{The overall PDF uncertainty of the inclusive
jet cross section is plotted as the two red curves.
The points are CDF data;
the error bars are statistical only and the systematic
shifts have not been subtracted.
\label{fig:CDFjetE}}
\end{figure}

Figure \ref{fig:CDFjetE} shows the overall PDF uncertainty
calculated from the ``Master Equation,'' asymmetrically for
positive and negative differences.
The ordinate is the fractional difference between the
extreme cross section and the central (CTEQ6.1M)
prediction.
The outer curves are the upper and lower bounds
of the uncertainty band.
The CDF data is superimposed.

\begin{figure}[ht]
\begin{center}
\includegraphics[scale=1]{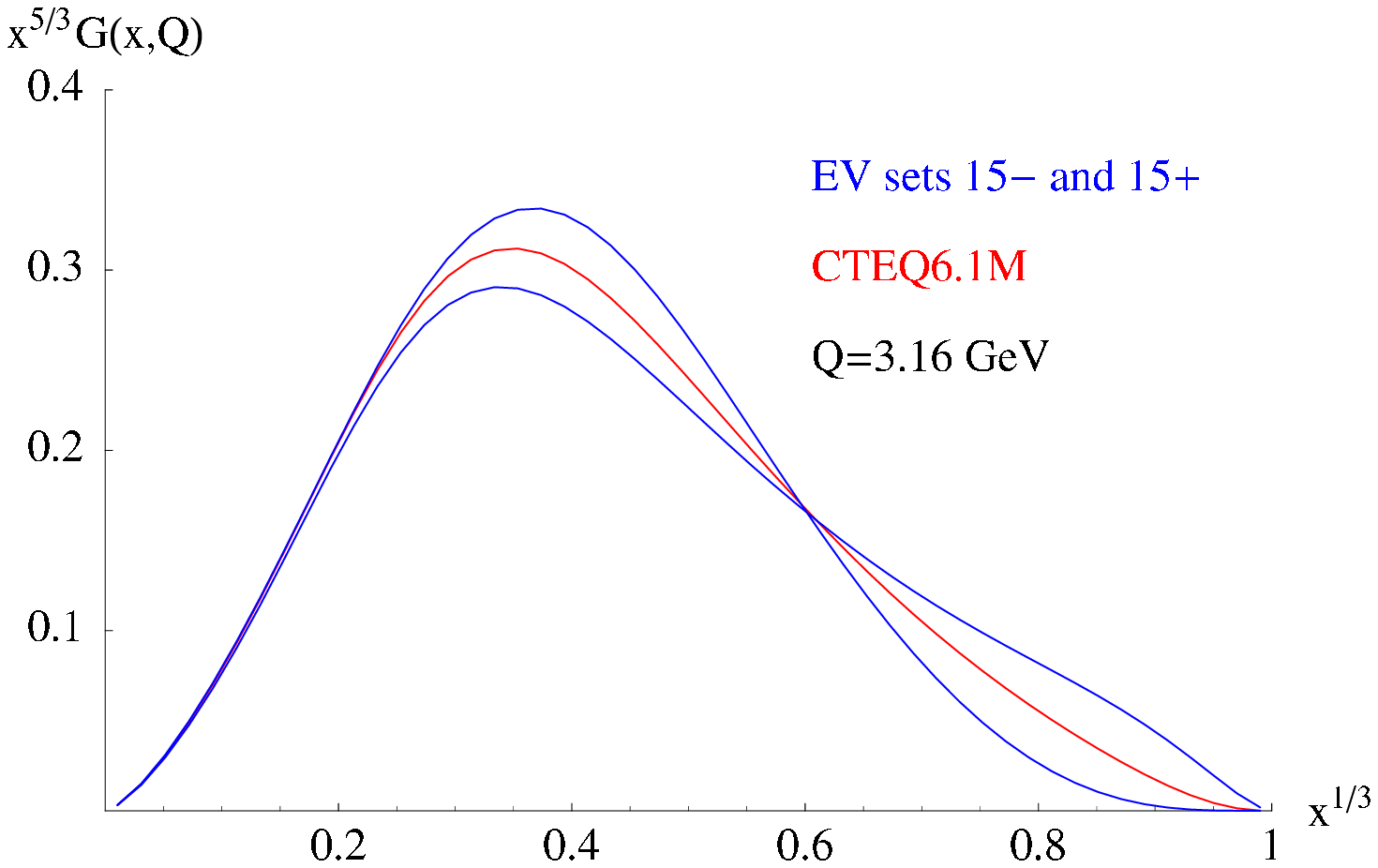}
\end{center}
\caption{Eigenvector 15.
Comparison between the gluon distributions for the
$+$ and $-$ displacements along eignevector 15
and the standard set CTEQ6.1M.
\label{fig:JPeig15}}
\end{figure}

The overall uncertainty in the calculation of
$(d\sigma/dE_{T})_{\rm CDF}$ is dominated by one of the
eigenvector directions.
The dominant eigenvector is ``eigenvector 15,'' i.e., that for
which the eigenvalue of the Hessian is 15th in order of magnitude.
The $+$ and $-$ displacements in parameter space along
this direction have a much larger effect on the value
of the cross section than any of the other eigenvector
directions.
In other words, $(d\sigma/dE_{T})_{\rm CDF}$ is most
sensitive to the parameter variations along this eigenvector.
Figure \ref{fig:JPeig15} shows the gluon distribution for
CTEQ6.1M and for the eigenvector basis sets in the $+$ and $-$
directions along eigenvector 15.
Eigenvector 15, like the 19 others, corresponds to some
linear combination of all of the 20 fitting parameters.
The exceptional feature of eigenvector 15 is that it
represents the largest change in the high-$x$ gluon behavior.

In the new CTEQ6.1M analysis, the eigenvector basis sets have
been carefully constructed to have
$\widehat{\chi}^{2}=\widehat{\chi}_{0}^{2}+100$.\footnote{%
The earlier CTEQ6 analysis~\cite{cteq6}
made an approximation of the NLO cross sections---holding
the $K$-factor tables for DIS and DY processes constant
as the PDF parameters varied in the neighborhood of the
minimum---such that $\Delta\widehat{\chi}^{2}$ was not
strictly constant at $T^{2}$ for the Hessian PDF sets.
The CTEQ6.1M analysis is preferred although the changes
from CTEQ6 are small.}
But $\widehat{\chi}^{2}$ is the {\em global} chi-square
function.
The criterion $\Delta\widehat{\chi}^{2}=100$ is only a reasonable
rule of thumb, and a more thorough analysis---inspecting the fit for
each data set---is needed to justify the extreme sets
$a_{15}^{+}$ and $a_{15}^{-}$.
We must ask whether these variations could be ruled out by any
specific experiment.

The theoretical cross section most affected by displacements
along eigenvector 15 is the inclusive jet cross section.
Table \ref{tab:ev15} shows the $\widehat{\chi}^{2}$ values
for the CDF and D\O\ data for
three PDF sets ($a_{0}$, $a_{15}^{+}$ and $a_{15}^{-}$.)
Because of the significant systematic uncertainties
in the experimental data, represented by the error
correlation matrices published by the collaborations,
we might conclude that the extreme sets $a_{15}^{+}$
and $a_{15}^{-}$ are still acceptable fits
to the data within the systematic errors.

\begin{table}[ht]
\begin{center}
\begin{tabular}{|c||c|c||c|c|}
\hline
PDF set & D\O\ $\chi^{2}$ & D\O\ $\chi^{2}/N$ 
& CDF $\chi^{2}$ & CDF $\chi^{2}/N$ \\
 & & $(N=90)$ & & $(N=33)$ \\
\hline
CTEQ6.1M & $52.26$ & $0.581$ & $57.67$ & $1.748$  \\
EV set $15+$ & $111.90$ & $1.243$ & $53.05$ & $1.608$ \\
EV set $15-$ & $39.17$ & $0.435$ & $67.35$ & $2.041$ \\ 
\hline
\end{tabular}
\end{center}
\caption{$\chi^{2}/N$ for the standard PDF's (CTEQ6.1M)
and the extreme variations in the $+$ and $-$ directions
along eigenvector 15.
\label{tab:ev15}}
\end{table}

As implied by the table above,
much of the increase in $\chi^{2}$
(50 out of 100) for eigenvector +15 comes from
the D\O\ jet cross section.
The Run 1b D\O\ cross section can tolerate some increase
compared to the CTEQ6.1M predictions, but the increase
corresponding to eigenvector +15 is in fact too extreme.
It is not surprising that such a large portion of
the increase in $\chi^{2}$ comes from the D\O\ jet data:
as shown in Fig.\,\ref{fig:JPeig15},
this eigenvector probes primarily the high-$x$
gluon distribution and the other data sets
(with the exception of the CDF jet data)
are fairly insensitive to this direction.
The opposite direction, eigenvector $-15$, gives
the lowest $\chi^{2}$ for the D\O\ jet data.
(Of course the $\chi^{2}$'s of other experiments
are larger so that the net change is $100$.)
The cross section in this case is less than that
for CTEQ6.1M, by as much as $10$ -- $20$\%.
The fact that $\widehat{\chi}^{2}/N$ is much less than 1
for the D\O\ data suggests that the size and correlation
of systematic errors are not accurately understood and
thus CTEQ6.1M gives a satisfactory fit although not the best fit.

\clearpage

\begin{figure}[ht]
\begin{center}
\includegraphics[scale=1]{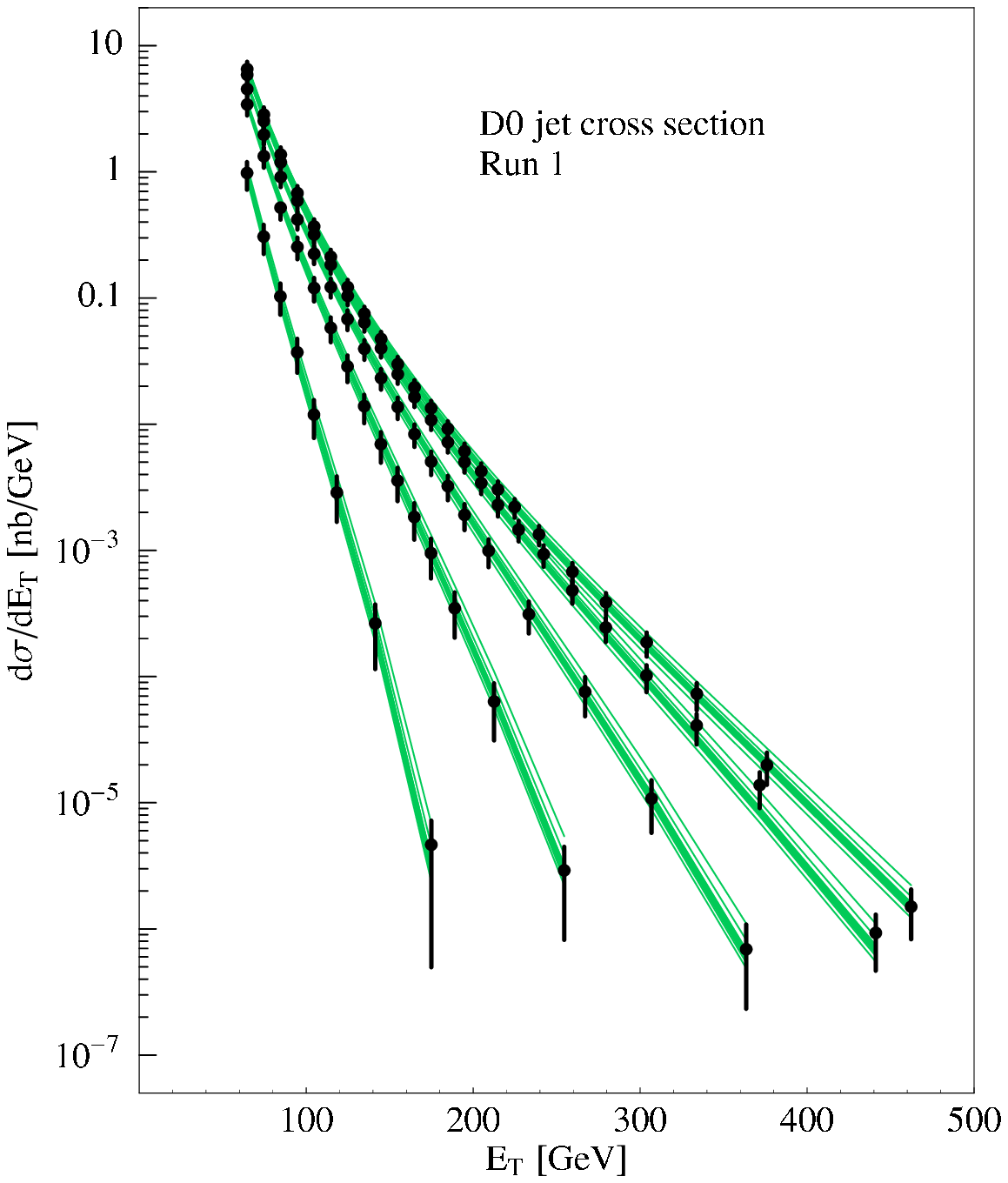}
\end{center}
\caption{The inclusive jet cross section as a function
of $E_{T}$ for the five rapidity bins of the Run 1b
D\O\ measurements.
Predictions of all 40 eigenvector basis sets are superimposed.
The points are the D\O\ data, and the error bars are the
statistical and systematic errors combined in quadrature.
\label{fig:D0jetA}}
\end{figure}

\begin{figure}[ht]
\begin{center}
\includegraphics[scale=1]{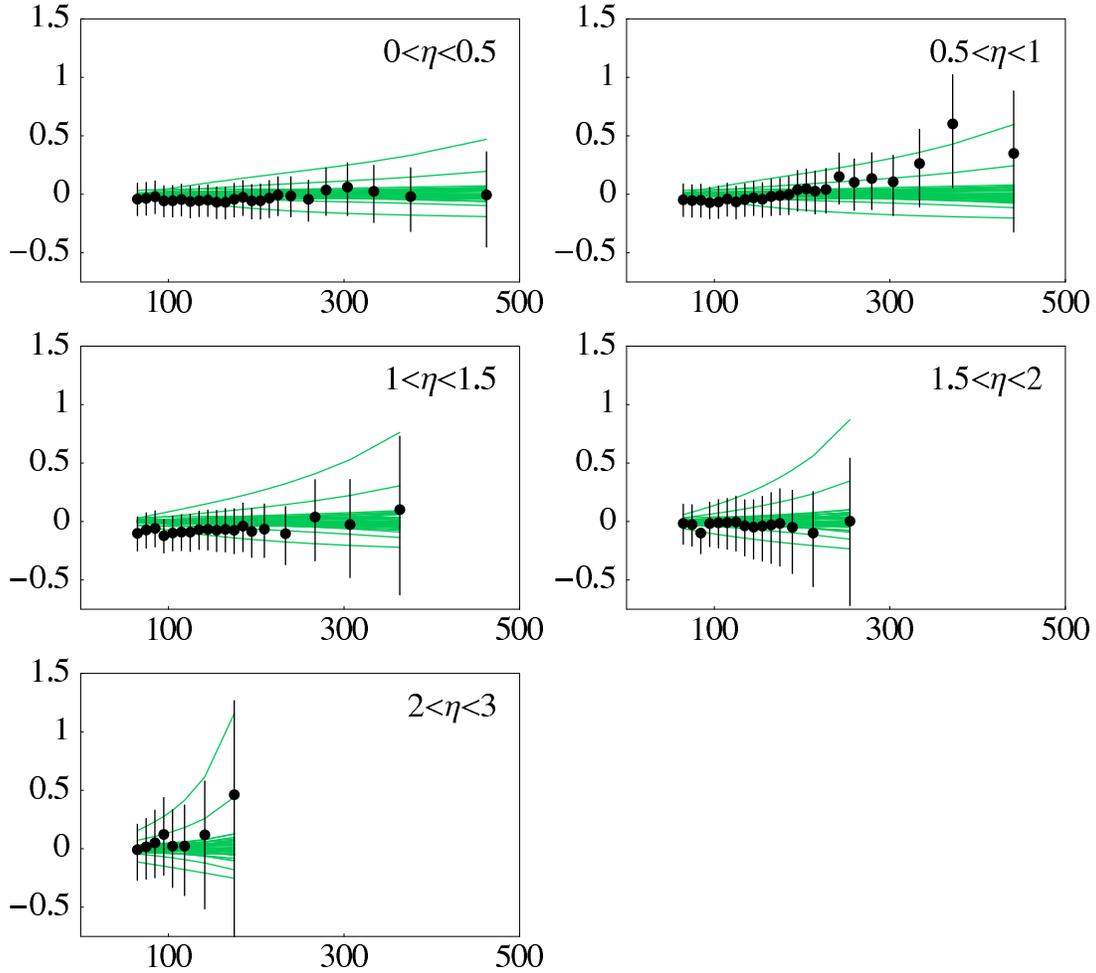}
\end{center}
\caption{The D\O\ inclusive jet cross section as a function
of $E_{T}$ for five rapidity bins, plotted as fractional differences
compared to the central (CTEQ6.1M) model.
The abscissa is $E_{T}$ in GeV and the ordinate
is $\left[T(a)-T(a_{0})\right]/T(a_{0})$ for the curves
and $(D-T)/T$ for the points.
\label{fig:D0jetB}}
\end{figure}

Figures for the D\O\ jet data analogous to Figs.\ \ref{fig:CDFjetA} 
and \ref{fig:CDFjetB} are shown
in Figs.\ \ref{fig:D0jetA}
and \ref{fig:D0jetB}.
Figure \ref{fig:D0jetA} shows the
inclusive jet cross section as a function of $E_{T}$
for the five rapidity bins of the D\O\ measurements.
Calculations for all 40 eigenvector basis sets are superimposed.
The points are the D\O\ data and the error bars equal
the single-point statistical and systematic errors added
in quadrature.\footnote{%
The systematic errors for the D\O\ data are supplied by the
collaboration only in the form of correlation matrices.
Unlike the CDF data, for which the standard deviations of
individual systematic errors have been published,
we are unable to separate the optimal systematic shifts
for subtraction from the D\O\ data.}
Figure \ref{fig:D0jetB} shows the inclusive jet cross section
as a function of $E_{T}$ for the five rapidity bins,
plotted as fractional differences compared to the central
(CTEQ6.1M) model.

\clearpage
\setcounter{section}{2}

\section{Reliability of next-to-leading-order predictions}
\label{sec:reliable}

Thus far, the discussion of uncertainties has dealt primarily with the 
issue of the propagation of experimental errors on the observables and 
their effect on the parameters of the parton distributions.
There are, of course, issues related to theoretical uncertainties
both in the global fit itself and in the theoretical calculations
for jet production.
However, it is far more difficult to quantify these.
We have previously described studies of some theoretical
uncertainties~\cite{cteq6}.
For example, we have chosen cuts on $Q$ and $W$ for the DIS
data sets used for CTEQ6 and earlier analyses in order to reduce
the theoretical uncertainties due to effects from higher twist
and related terms.
Also, we have varied the functional form of the parametrization
to ensure that sufficient flexibility to cover the wide kinematic
range of $x$ and $Q$ is available.
In this section we concentrate on the theoretical uncertainties
related to NLO calculations for jet production.
The response of the predictions to variations
in the renormalization and factorizations scales yields 
some information on the theoretical uncertainties.
In this section the scale dependence of the jet cross 
sections and the effects on the fits of varying the scales will be discussed. 
Some estimates of the possible effects of two-loop corrections will also be 
presented. 

\begin{figure}[ht]
 \includegraphics[width=0.8\textwidth]{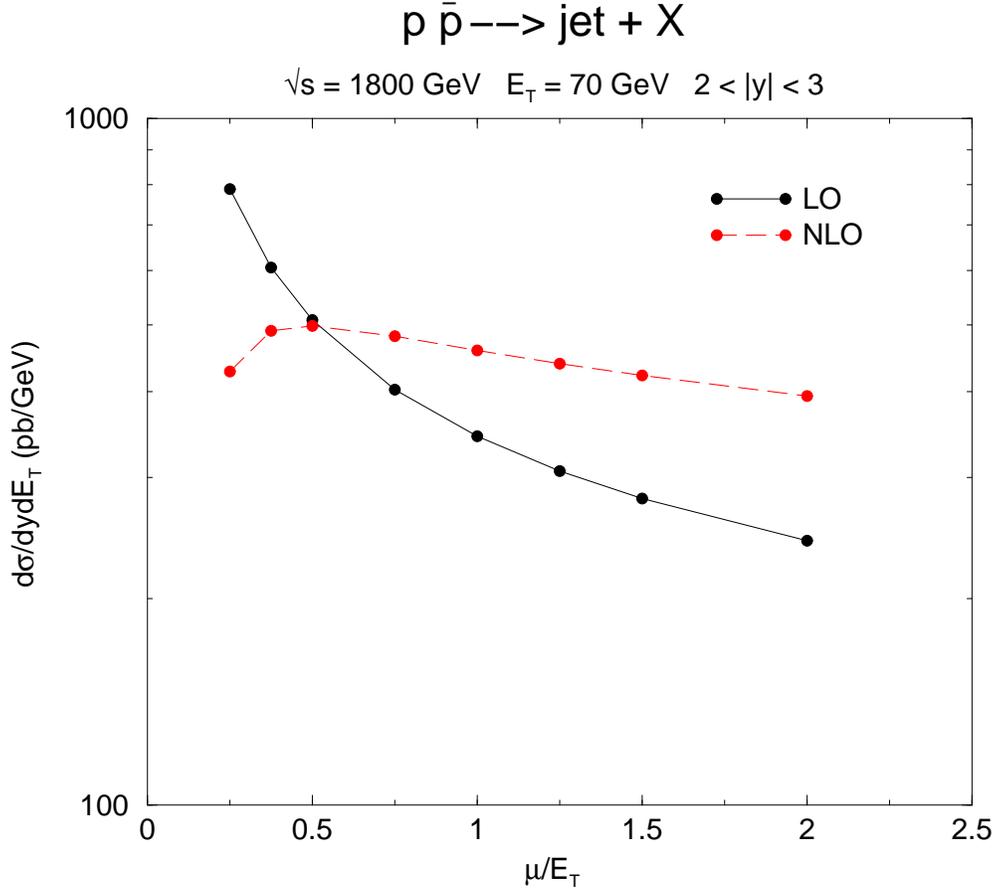}
 \caption{Comparison of the scale dependence of the jet cross section at 
 $E_T$=70 GeV/c for the leading- and next-to-leading-logarithm 
 calculations}
 \label{scale_70}
\end{figure}

\begin{figure}[ht]
 \includegraphics[width=0.8\textwidth]{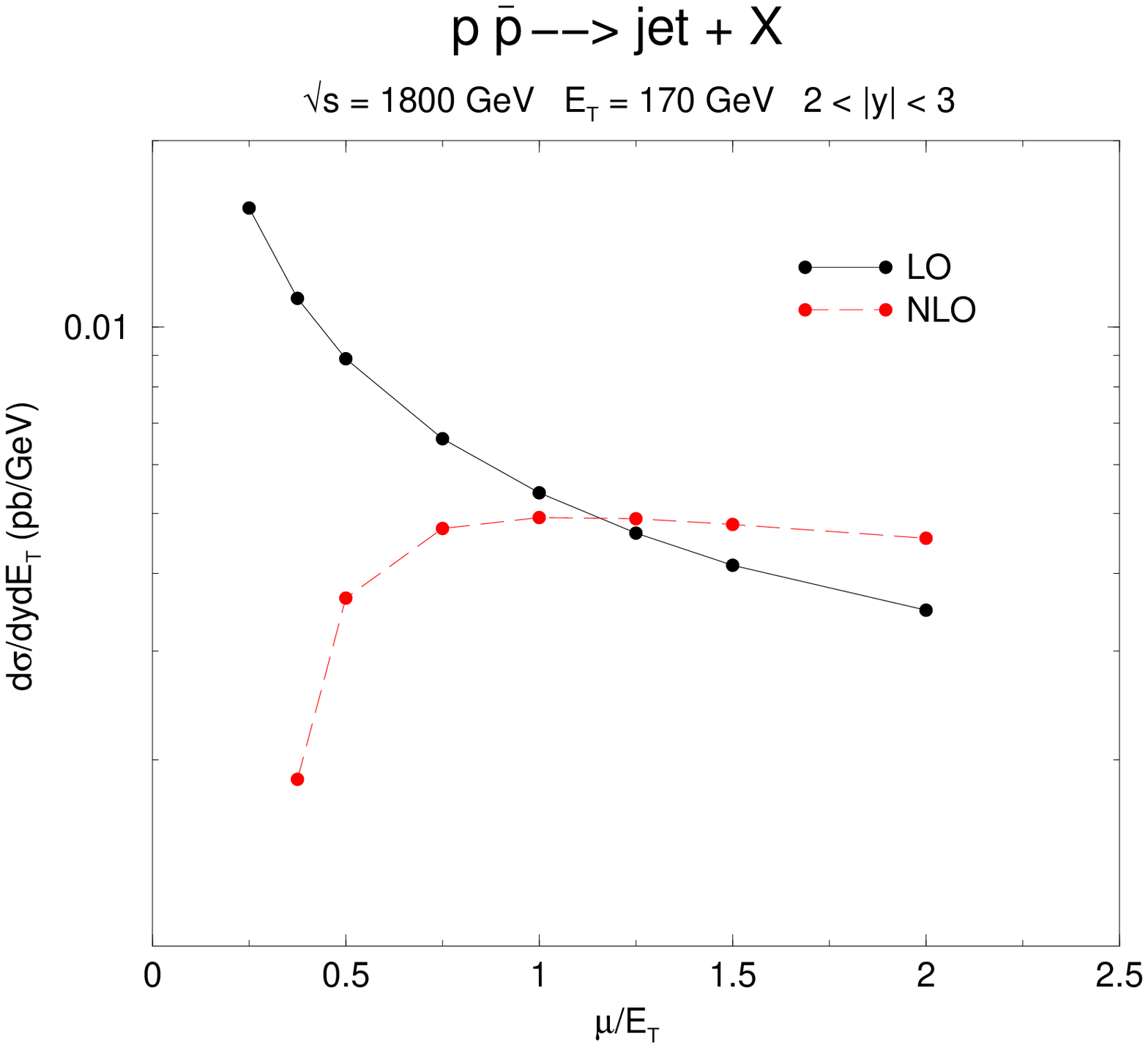}
 \caption{Comparison of the scale dependence of the jet cross section at 
 $E_T$=170 GeV/c for the leading- and next-to-leading-logarithm 
 calculations}
 \label{scale_170}
\end{figure}

\subsection{Scale dependence of the jet cross section}
\label{sec:scaledep}

Leading-logarithm calculations of the jet cross section at high-$E_T$  
generally decrease monotonically as the factorization and renormalization 
scales are increased. The scale variation of next-to-leading-logarithm 
calculations is generally reduced as compared to that of the leading-log 
calculations.  A brief review of how this comes about is 
presented in the Appendix.

The increased rapidity coverage available with the recent D\O\ data 
\cite{d01b} has provided additional constraints on the gluon distribution. 
However, if the theoretical uncertainties become large at high rapidity, 
then these constraints may not be as useful as first believed.
As one approaches the edge of phase space,
it might be anticipated that there will be 
large logarithms due to the constraints on multiple gluon emission.
The effects of these large logs will be investigated in 
Sec.\ (\ref{sec:resum}).
Their effects may show up already in the next-to-leading-log
calculations used in the global fits, so it is important 
to look for any anomalous behavior in the predictions. 

The first step is to examine the scale dependence of the jet cross section 
predictions in the highest rapidity bin of the D\O\ data.
For ease of 
presentation, the factorization and renormalization scales have been 
chosen to be the same.
In the jet cross section calculations used for the CTEQ PDF fits,
and in this paper, a scale $\mu_R = \mu_F = E_T/2$ has been used.
Fig.\,\ref{scale_70} shows the comparison of the 
scale dependences of the leading-log and next-to-leading-log calculations 
at $E_T=70$\,GeV/$c$.
The monotonically decreasing behavior of the leading-log 
results is evident, as is the roughly quadratic dependence of the 
next-to-leading-log curve. Both behaviors are consistent with the 
expectations  discussed in the Appendix. Similar results are shown in 
Fig.\,\ref{scale_170} for $E_T=170$\,GeV/$c$.
Both figures show that the scale dependence of the
next-to-leading-log calculation is greatly reduced 
with respect to that of the leading-log calculation,
provided that one avoids scale choices much smaller that $\mu = E_T$. 

It is interesting to note that the $K$-factor (NLO/LO) for $\mu = E_T/2$ 
is approximately unity for $E_T = 70$\,GeV/$c$,
while it is significantly less for $E_T = 170$\,GeV/$c$.
The reasons for this are discussed in more detail in Appendix A. 

\subsection{Fits with different scales}
\label{sec:fit_scales}

The next point to be addressed is the effect on the fits of varying the 
choice of scale. To investigate this point, the CTEQ6M 
fits\footnote{%
This procedure has not been repeated with the CTEQ6.1M PDF's.
However, since the changes in the region of the D\O\ and CDF
data were minimal, the scale dependence in the 6M and 6.1M sets
should be essentially the same.}
were repeated with scale choices of $\mu = E_T$ and $2 E_T$.
As compared to the original fit with $\mu = E_T/2$
the latter two yielded increases in chi-square of
16 and 70, respectively.
Thus, the choice of $E_T/2$ used in the CTEQ6M fits yields the
best chi-square of the three choices, but this does not constitute
a fit for the optimal scale.
Note, however, that the rapid increase in the scale dependence
for choices below about $E_T/2$ shown in Figs.\,\ref{scale_70}
and \ref{scale_170} precludes fits with significantly
smaller scales while the larger scale choices
show an increase in chi-square.

\begin{figure}[ht]
\begin{center}
\includegraphics[width=0.8\textwidth]{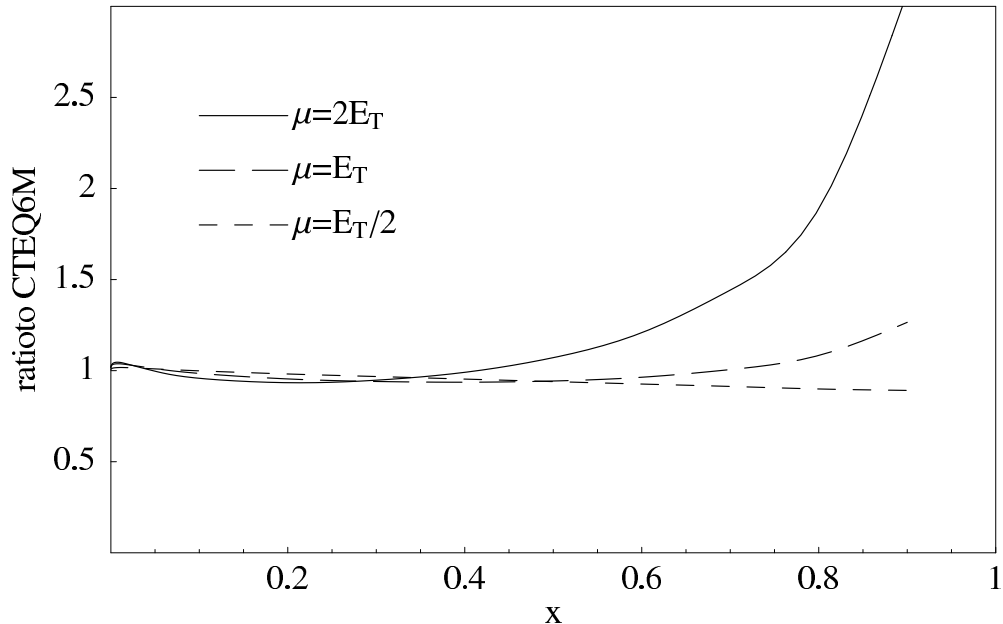}
\end{center}
\caption{Comparison of fitted gluon distributions for fits made with 
different scale choices for the jet cross section.
Comparison to Fig.\,\ref{fig:PDFunc} shows that this
theoretical error is smaller than the uncertainty due to
experimental errors.
\label{fig:glu_ratio}}
\end{figure}

The gluon distributions from these fits are shown
in Fig.\,\ref{fig:glu_ratio} relative to the standard CTEQ6M
distribution corresponding to $\mu = E_T/2$.
The effects at values of $x$ below about $0.4$ are minimal,
but the effects at larger $x$ values are significant.
In particular, the reduction in the cross section due
to the larger scale choice of $2E_T$ is compensated
by an increase of the gluon distribution.
However, it should be noted that the chi-square increased
by 70 units for this fit.
This is still less than the chi-square increase of 100 that
determines the 40 error PDF's.

A comparison of Fig.\ \ref{fig:glu_ratio}
to Fig.\ \ref{fig:PDFunc} indicates that
the uncertainty on the gluon distribution due to the scale choice
for jet production in the global fits is everywhere less than the
uncertainty from the treatment of the experimental errors. 

It is important to bear in mind that the gluon distribution itself is not an 
observable quantity. Rather, the gluon and quark distributions contribute via 
convolutions with appropriate hard scattering cross sections. 
Accordingly, the impression provided by Fig.\ \ref{fig:glu_ratio} may be 
somewhat misleading as much of the gluon variation is offset by changes in 
the hard scattering cross section with which it is convoluted, with the end 
result being that the variation in the physical observables is much less.

\begin{figure}[ht]
\begin{center}
\includegraphics[width=0.6\textwidth]{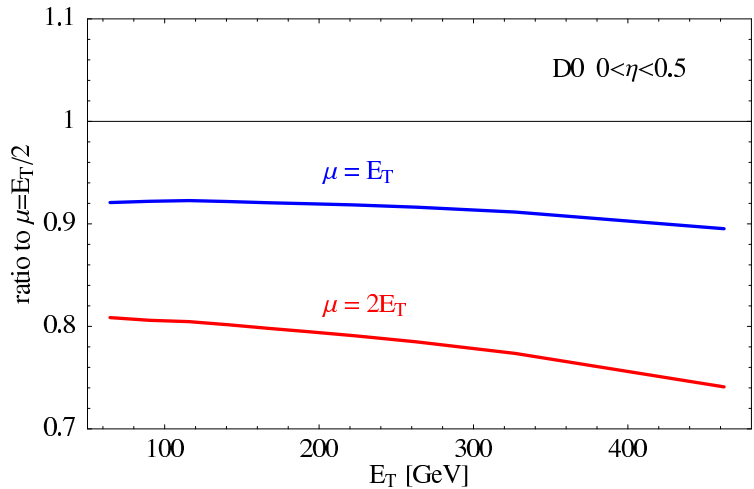}
\end{center}
\caption{Scale dependence of the jet cross section prediction
for $0<y<0.5$.
\label{fig:CSscaledep0}}
\end{figure}
\begin{figure}[ht]
\begin{center}
\includegraphics[width=0.6\textwidth]{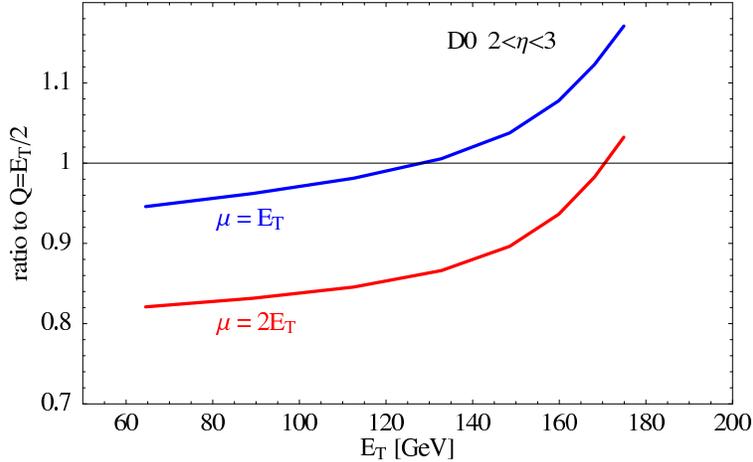}
\end{center}
\caption{Scale dependence of the jet cross section prediction
for $2<y<3$.
\label{fig:CSscaledep2}}
\end{figure}

The jet cross section predictions using scales of $E_T/2$,
$E_T$ and $2E_T$ are shown in Figs.\,\ref{fig:CSscaledep0}
and \ref{fig:CSscaledep2} for the
D\O\ rapidity regions of $0.0$ -- $0.5$ and $2.0$ -- $3.0$.
These curves may be compared to those for bins 1 and 5 in Fig.\, 
\ref{fig:D0jetB}. The scale dependence variations are generally within 
the bands formed by the 40 eigenvector basis sets, indicating that the 
uncertainty introduced by the scale dependence of the jet cross sections 
is less than or on the order of that coming from the experimental errors.



\subsection{Threshold resummation}
\label{sec:resum}

\begin{figure}[ht]
\begin{center}
\includegraphics[height=4 in, angle=0]{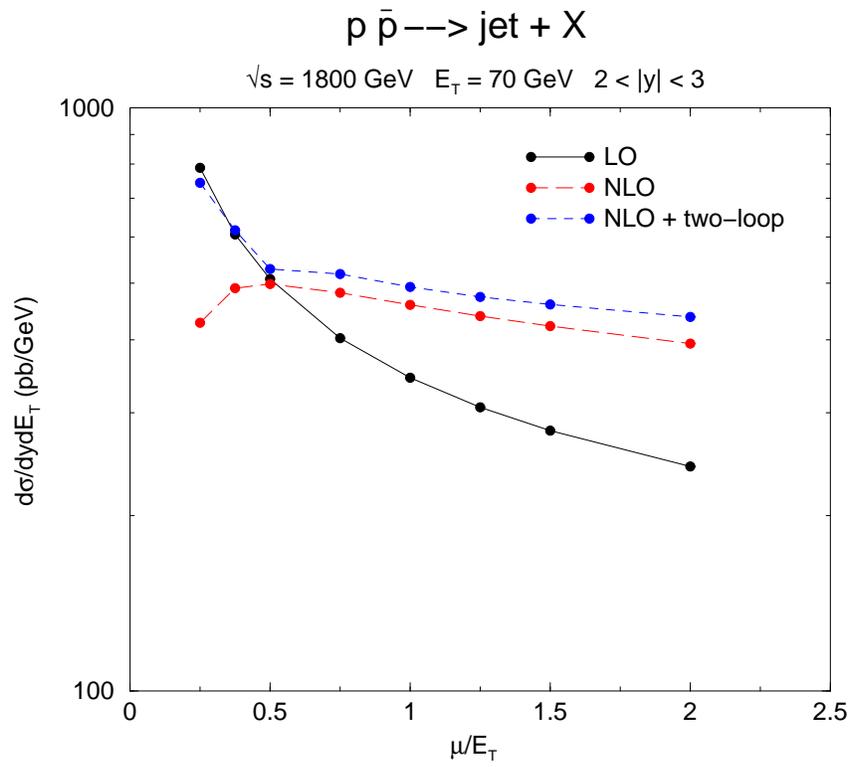}
\end{center}
\caption{Same as Fig.\ \ref{scale_70} with the addition of 
the estimated two-loop contribution expected from threshold 
resummation.
\label{scale_twoloop_70}}
\end{figure}

\begin{figure}[ht]
\begin{center}
\includegraphics[height=4 in, angle=0]{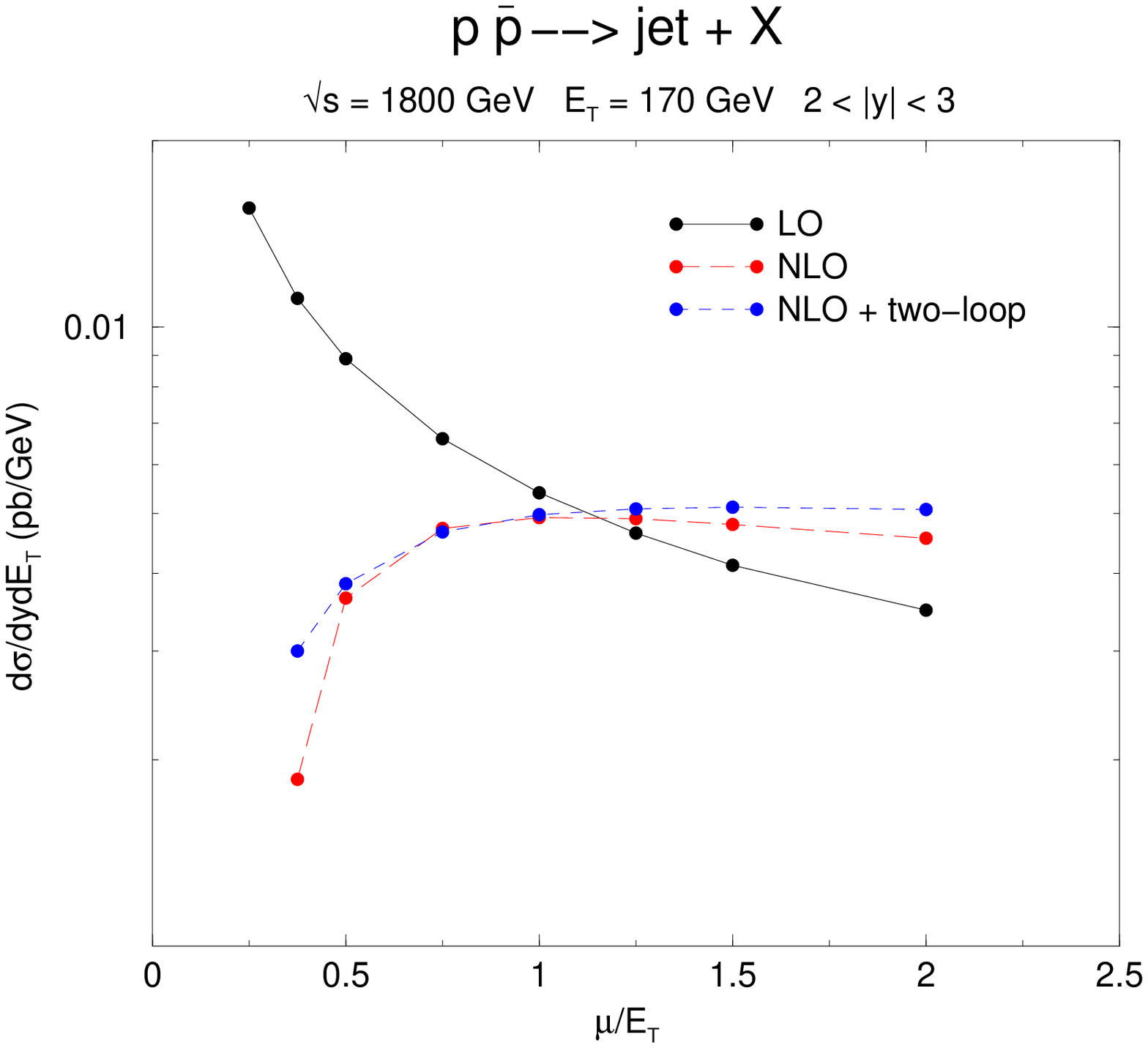}
\end{center}
\caption{Same as Fig.\ (\ref{scale_170}) with the addition of 
the estimated two-loop contribution expected from threshold 
resummation.
\label{scale_twoloop_170}}
\end{figure}

As noted in the introduction to this section,
there are potentially significant contributions to the
jet cross resulting from phase space limitations on gluon
emission as one goes to large $E_T$ and large rapidity. 
Such corrections can be treated
by threshold resummation techniques~\cite{resum}.
In Ref.\,\cite{twoloop} these corrections for 
jet production were estimated at the two-loop level for central rapidity 
values and found to be small.
The calculation has been repeated for the 
largest rapidity bin covered in the D\O\ data~\cite{d01b}
and the results are shown in Figs.\,\ref{scale_twoloop_70}
and \ref{scale_twoloop_170} by the dashed lines.
As was the case in the central region,
the two-loop estimates of the next-to-next-to-leading-logarithm 
contributions are not large for the scale choices considered
in this analysis.

To summarize the results of this section,
the next-to-leading-logarithm jet calculations show the expected
reduction in scale dependence, even for large values of $E_T$
in the highest rapidity bin in the existing data set. 
Nevertheless, the remaining scale variation does show up
in an increased uncertainty in the gluon distribution at
large values of $x$.
This uncertainty is, of course, partially offset by the
compensating scale dependence of the hard scattering cross section.
Even so, the scale dependence of the next-to-leading-log calculations
remains as a contributing factor for the overall uncertainty of the
predictions for new processes.
Finally, the next-to-next-to-leading-log corrections estimated
by expanding the threshold resummation predictions for jet production
remain small, even at high $E_T$ in the highest rapidity bin considered,
lending support to the use of the next-to-leading-log expressions
in the global fit.


\clearpage
\setcounter{section}{3}
\section{Room for New Physics in Run 1b}

Hadron collider jet data have traditionally been used to constrain
models of new physics, usually models of quark substructure.
The measurements include dijet angular distributions that are
insensitive to theoretical uncertainties due to parton
distributions~\cite{angdist}.
Inclusive jet cross sections or dijet mass distributions
typically provide better probes of quark compositeness,
but are sensitive to parton distributions at large $x$
which have large uncertainties~\cite{dijetmass}.
Currently the best limits are provided by a hybrid approach
that uses the ratio of dijets in different rapidity bins~\cite{d0dijet}.
This approach has the advantage that many theoretical and experimental
systematic errors cancel out in the ratio.
It has the disadvantage that it removes some useful information,
such as the absolute values ot the cross sections in the two
rapidity intervals.

There are several reasons to revisit this issue.
First, the published limits are based on parton distributions
several generations old, and the best-fit gluon distribution
at large $x$ is significantly changed.
Second, the recent jet measurements at large rapidity provide
constraints on the large $x$ partons while being largely
insensitive to new physics.
Finally, and most important, the global analysis machinery
has never been used in these analyses to refit the parton
distributions in the presence of a new component to the QCD calculation.
If the constraints on the parton distributions are not sufficient,
it is possible that new physics could hide inside the PDF uncertainties,
or current limits may be weakened.

Compositeness signals are typically parameterized by a mass scale
$\Lambda$ which characterizes the quark substructure coupling.
The coupling is approximated by a four-Fermi contact interaction
giving rise to an effective Lagrangian.
Only the term describing left-handed coupling of quarks and
anti-quarks has been calculated~\cite{comp},
\begin{equation}\label{eq:SKmodel}
\mathcal{L}_{qq}
= A (2\pi/\Lambda^{2})
(\overline{q}_{L}\gamma^{\mu}q_{L})
(\overline{q}_{L}\gamma_{\mu}q_{L})
\qquad \mbox{where} \qquad A=\pm1.
\end{equation}

The simplest squared amplitude
is the quark anti-quark s-channel annihilation process,
with the first term the normal QCD interaction and the
second due to substructure:
\begin{equation}
\mid A(u\overline{u}\rightarrow d\overline{d})\mid^{2}
=\frac{4}{9}\alpha_{S}^{2}(Q^{2})
\frac{(\widehat{u}^{2}+\widehat{t}^{2})}{\widehat{s}^{2}}
+\frac{A\widehat{u}}{\Lambda^{2}}.
\end{equation}
Since $\widehat{u}$ is negative, constructive
(destructive) interference in
this process occurs when $A=-1 \ (+1)$.
This is generally true for the other scattering terms as well,  
which will be included in the calculations.
As in the experimental papers, we will use the
quark compositeness implementation in PYTHIA~\cite{pythia}
to form the ratio of QCD+substructure/QCD, then multiply this
ratio by the NLO QCD calculation for each iteration of parton
distribution as the PDF parameters are being fit.  As expected,  
we find larger cross sections with the constructive interference 
choice $A=-1$,  approximately a factor of 10 larger.
However, the ratio of the first two eta bins used by D\O\ 
is slightly larger with the destructive interference 
term $A=1$, and this is what is used for the current
best limit of $\Lambda > 2.7$\,TeV~\cite{d0dijet}.  
Therefore we have  used the destructive interference
choice for illustration in the rest of this paper.

\begin{figure}[ht]
\begin{center}
\includegraphics[width=0.8\textwidth]
{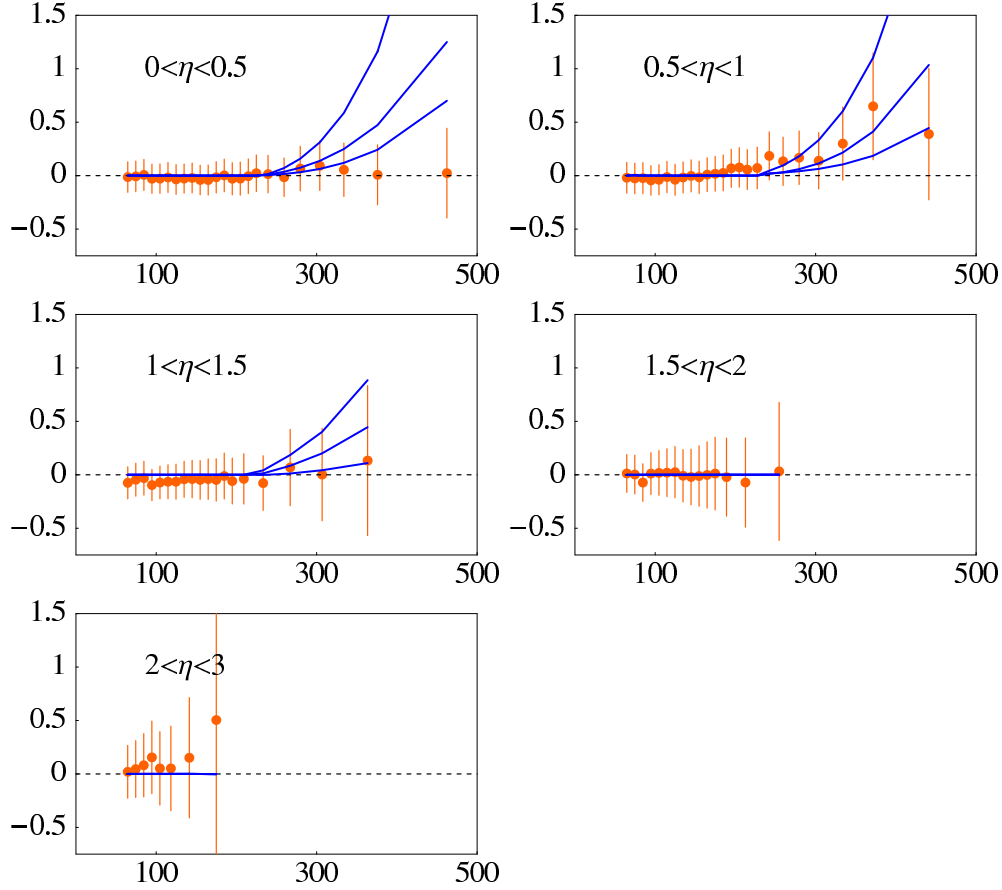}
\end{center}
\caption{Comparison of three models in which a hypothetical
contact interaction contributes to jet production.
The ordinate is the fractional difference between the model
and the standard theory;
the abscissa is the jet $E_{T}$ in GeV.
The contact interaction has $\Lambda=1.6$, $2.0$, and $2.4$\,TeV
for the three curves.
In all cases the PDF's are the standard CTEQ6.1M.
The D\O\ data are superimposed for comparison.
\label{fig:ContactInt}}
\end{figure}

Figure \ref{fig:ContactInt} shows the cross section for jet
production including a contact interaction, for three values
of $\Lambda$: $1.6$, $2.0$ and $2.4$\,TeV.
The cross section is plotted as a ratio to the pure
QCD prediction with the CTEQ6.1M parton distributions.
The range of kinematic parameters are the same as the
D\O\ data,
and the D\O\ data is superimposed on the plots with error
bars of statistical and systematic errors combined in quadrature.

If the effect of the contact interaction is smaller than the
PDF uncertainty of the QCD prediction then the data cannot
rule out the compositeness model; any difference between theory
and data could be explained either as a PDF effect or as
a sign of new physics.
However, if the compositeness model disagrees with the
data by an amount that is larger than the PDF uncertainty,
the model is ruled out.
The PDF uncertainty of the QCD prediction is comparable to the
experimental error bars (see Fig.\ \ref{fig:D0jetB}), so we may
use those error bars to judge whether the effect of the contact
interaction is smaller than the PDF uncertainty.
The contact interaction is seen to be ruled out
for $\Lambda=1.6$\,TeV, primarily from the disagreement
with the central rapidity interval,
$0<\eta<0.5$.
The cross section at large rapidity ($\eta>1.5$) is
not sensitive to the contact interaction.

The largest two $\Lambda$ values, $\Lambda=2.0, 2.4$\,TeV,
are not inconsistent with the D\O\ data on inclusive jet
production because the effect of the contact interaction
is within the PDF uncertainty.
The global fit for $\Lambda=2.4$\,TeV is even slightly
better than for the pure QCD model, although the difference
in the overall quality of the fit
($\Delta\chi^{2}_{\rm global}=9$)
is much less than our standard tolerance for PDF uncertainty
($\Delta\chi^{2}_{\rm global} \lesssim 100$).

\begin{figure}[ht]
\begin{center}
\includegraphics[width=0.8\textwidth]
{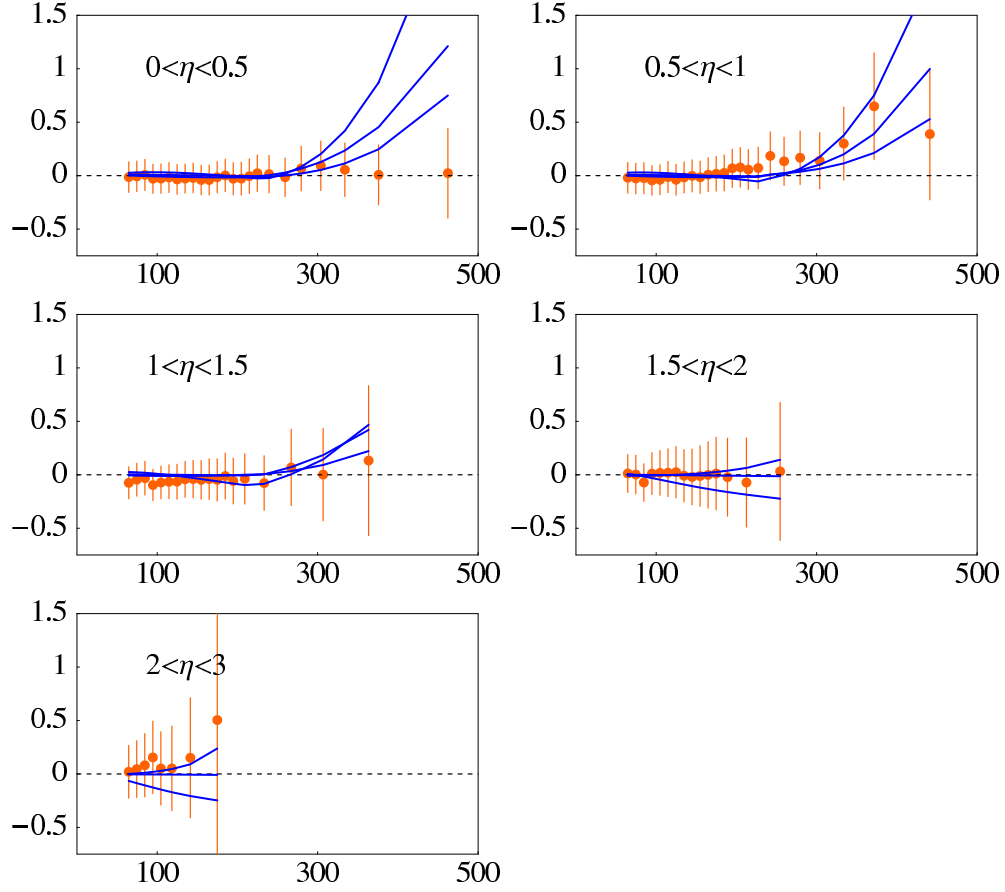}
\end{center}
\caption{Comparison of
three models in which a hypothetical contact interaction
contributes to jet production, with refitting of the PDF's.
The curves have the same meaning as in
Fig.\,\ref{fig:ContactInt}.
The PDF's for the three models have been refit to the
global data set to give the best fit including the contact
interaction in jet production.
\label{fig:ContactRefit}}
\end{figure}

If there is a contact interaction, then the CTEQ6.1M PDF's
would be inaccurate because the data were fit to a
pure QCD model.
Therefore a complete analysis of the contact interaction
requires that the PDF's be refit, comparing the data to
the theory with compositeness.
Figure \ref{fig:ContactRefit} shows the final result
of this analysis.
The cross section for jet production with the D\O\
kinematic parameters is again shown for three values of the
$\Lambda$ parameter ($1.6$, $2.0$ and $2.4$\, TeV)
as a ratio to the pure QCD model with CTEQ6.1M.
In Fig.\ \ref{fig:ContactRefit}
the PDF's for each $\Lambda$ value come from
a global fit that includes the contact interaction
in the theory of the jet cross section.
The smallest value of $\Lambda$ is now more
consistent with the data, but still ruled out.

The limit derived from this analysis,  $\Lambda>1.6$ TeV,  is
not as strong as those from either the CDF or D\O\ analyses that
use angular distribution information~\cite{angdist}.  This is a result
of the remaining uncertainty in the PDFs,  and the fact that
in this model of new physics the angular distributions are 
quite different than QCD.  Other deviations from the Standard Model
might change the absolute cross sections while maintaining a 
more QCD-like angular distribution.   In this case this 
analysis is the most in-depth study so far of allowed 
deviations in jet cross sections due to new physics. 

The limit derived from this analysis is also less than
that obtained by the D\O\ collaboration when using the
ratio of the dijet cross sections in the two rapidity bins
($0.0 - 0.5$ and $0.5 - 1.0$).
However, it should be noted that at least part of the
higher limit from D\O\ may be due to the dijet cross
section for $0.5 < \eta < 1.0$ being somewhat larger
than the standard model prediction, while the dijet cross
section for $0.0 < \eta < 0.5$ is in better agreement
with the SM prediction.
The larger the ratio of the $0.5 - 1.0$ to $0.0 - 0.5$
rapidity intervals, the stronger the compositeness limit.
The detailed information about the absolute values
of the cross sections in each rapidity bin
has not been used.

Table \ref{tbl:compositeness} shows how the $\chi^{2}$'s
are affected by the contact interaction.

\begin{table}
\begin{center}
\begin{tabular}{|c|c|c|c|c|} \hline
$\Lambda$\,[TeV] & PDF set & global $\chi^{2}$ 
 & D\O\ $\chi^{2}/N$ & CDF $\chi^{2}/N$
\\ \hline
1.6 & CTEQ6.1M & 2111 & 0.939 & 5.45 \\
2.0 & CTEQ6.1M & 1966 & 0.566 & 2.07 \\
2.4 & CTEQ6.1M & 1948 & 0.493 & 1.73 \\
$\infty$ & CTEQ6.1M & 1957 & 0.590 & 1.68 \\
\hline
1.6 & best refit & 2079 & 1.230 & 3.63 \\
2.0 & best refit & 1965 & 0.576 & 2.00 \\
2.4 & best refit & 1947 & 0.493 & 1.73 \\ 
\hline
\end{tabular}
\end{center}
\caption{$\chi^{2}$'s for three theoretical models with
a contact interaction.
\label{tbl:compositeness}}
\end{table}

\clearpage
\setcounter{section}{4}
\section{Predictions for Run 2}

The increase in the center-of-mass energy (from 1.8 to 1.96 TeV)
and the increased luminosity expected in Run 2 (from approximately
$100$\,pb$^{-1}$ to $2$\,--\,$15$\,fb$^{-1}$ will result in a dramatically
larger kinematic range for measuring jet production,
and thus for searching for possible new physics as well.
Also, in Run 2, the new CDF endplug calorimeter will allow
for an extension of measurements of the inclusive jet cross
section to the forward rapidity region, similar to the analysis
that has been performed by D\O\ in Run 1b. 

\begin{figure}[ht]
\begin{center}
\parbox{0.49\textwidth}
{\includegraphics[width=0.49\textwidth]{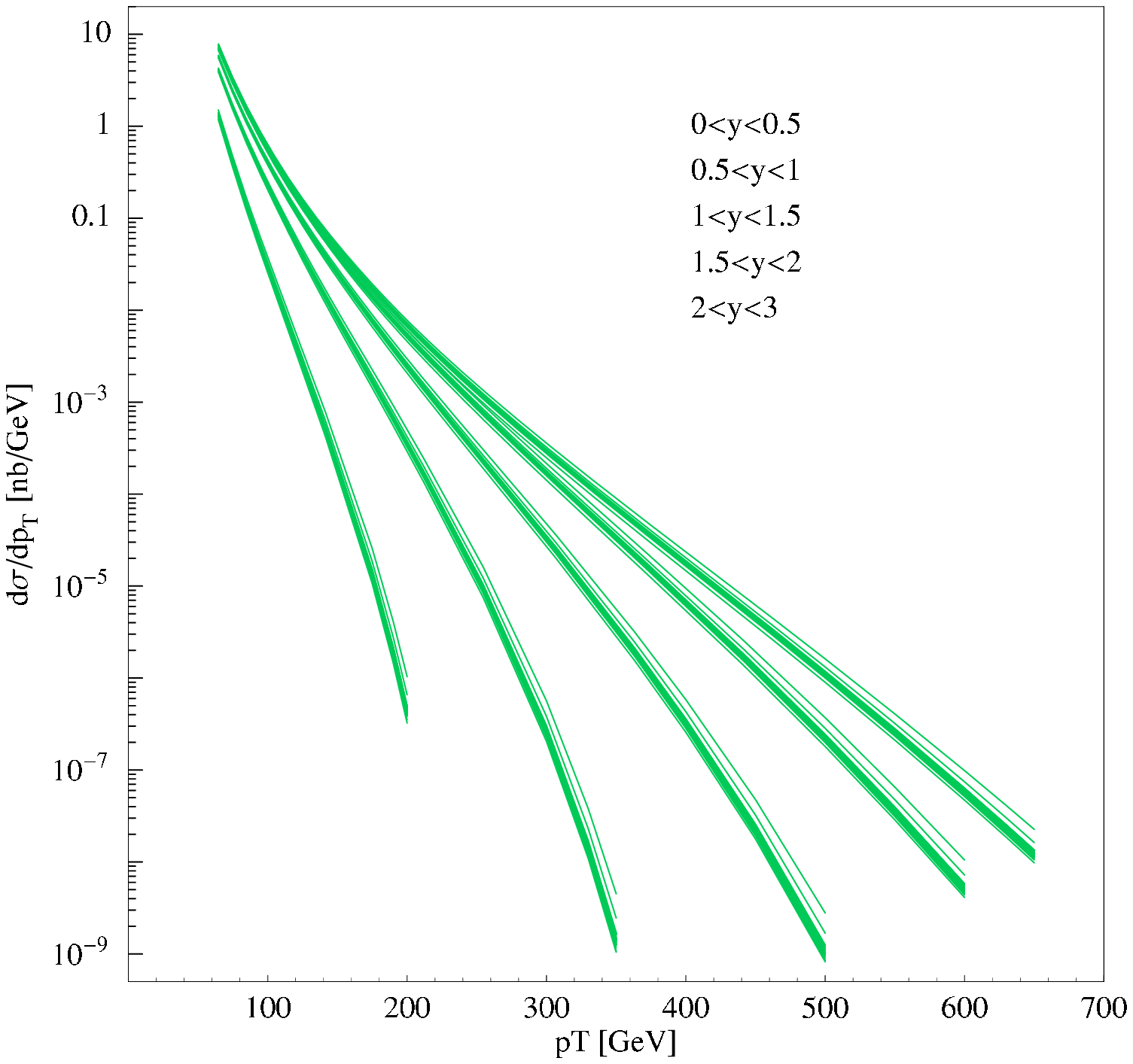}}
\hfill
\parbox{0.49\textwidth}
{\includegraphics[width=0.49\textwidth]{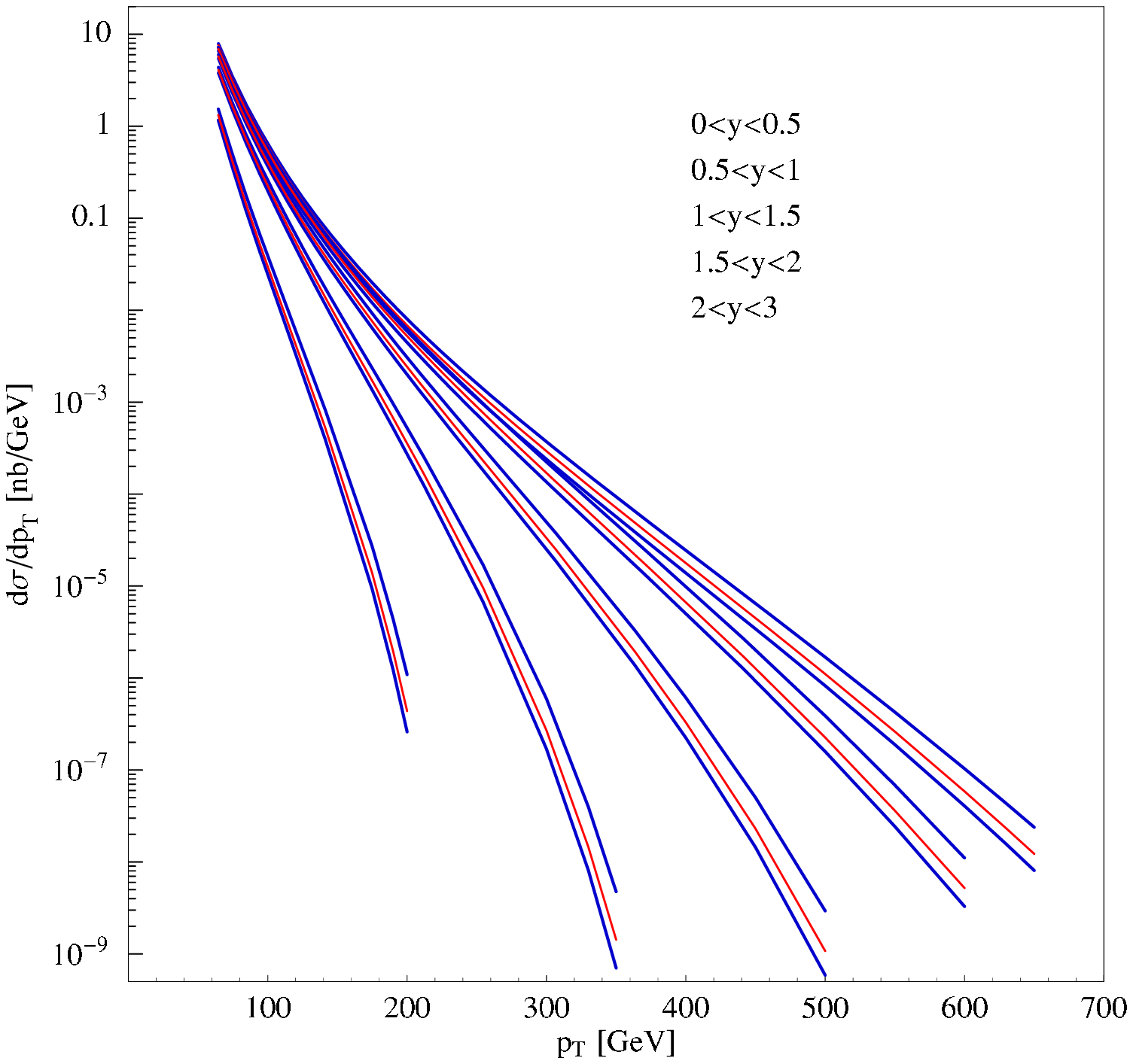}}
\end{center}
\caption{Run 2 cross section for the D\O\ rapidity bins.
Left: The 40 eigenvector basis sets.
Right: The central prediction and overall uncertainty band.
\label{fig:D0Run2}}
\end{figure}

\begin{figure}[ht]
\begin{center}
\includegraphics[width=0.79\textwidth]{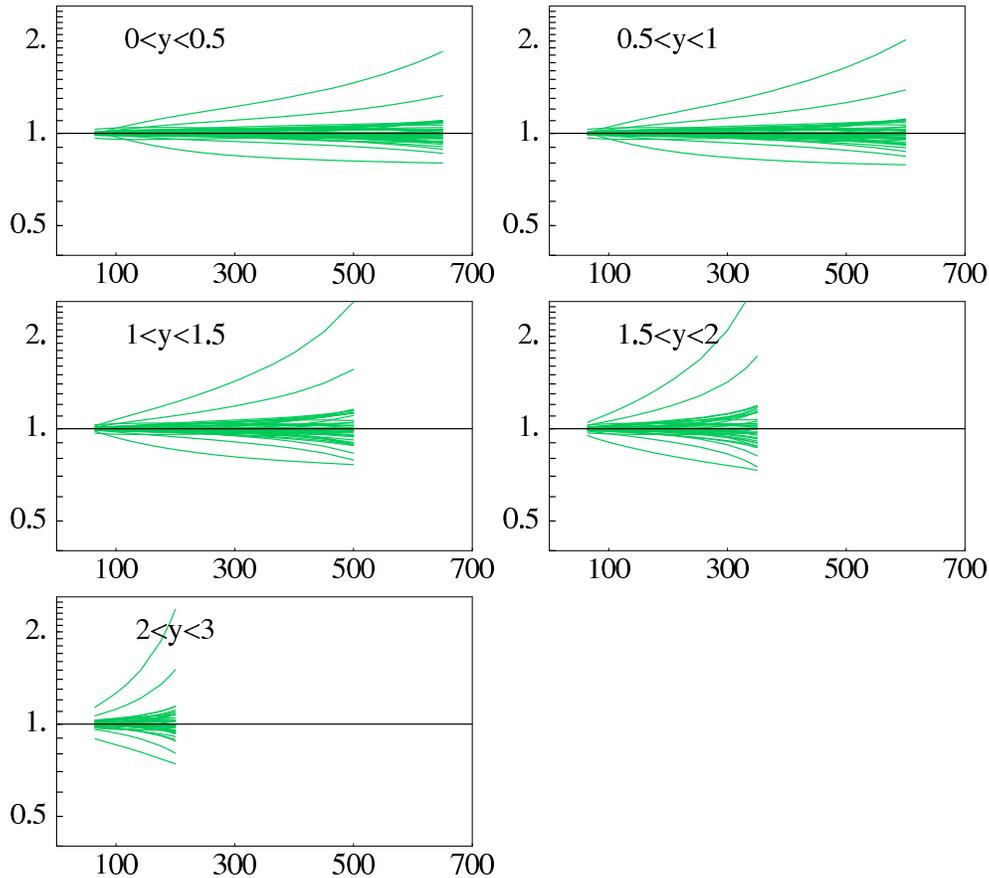}
\end{center}
\caption{Uncertainty range of the Run 2 cross section
for the D\O\ rapidity bins.
The curves show the ratios of the 40 eigenvector basis sets
to the central (CTEQ6.1M) prediction (ordinate)
versus $p_{T}$ in GeV (abscissa).
\label{fig:D0Run2ratio}}
\end{figure}

The predictions for the Run 2 jet cross section as a function
of transverse momentum (using the D\O\ rapidity intervals)
are shown in Fig.\,\ref{fig:D0Run2}.
With $2$\,fb$^{-1}$ (Run 2a) a measurement of the inclusive
jet cross section out to a transverse energy of
$600$\,GeV is possible,
while with $15$\,fb$^{-1}$ (Run 2b)
the measurement can be extended to almost $650$\,GeV.
The predictions are displayed on the left graph
in the form of the curves for the 40 eigenvector
basis sets of CTEQ6.1M.
The central prediction and overall uncertainty band
from the Master Equation (\ref{eq:MEasy}) are shown
separately on the right graph.
Figure \ref{fig:D0Run2ratio} shows the uncertainty range
of the Run 2 cross section by displaying the ratios of
the eigenvector basis sets to the central prediction.
Near the kinematic limit, the PDF uncertainties allow
the cross section predictions to be as much as a factor
of 2 larger than those from the central fit (CTEQ6.1M),
again due to eigenvector 15.

For the Run 2 jet cross section predictions, we have used the
Run 2 cone jet algorithm~\cite{run2work, ellis} which uses
4-vector kinematics both to define the jet and to specify its
transverse momentum and rapidity.
This algorithm has been adopted by both the CDF
and D\O\ collaborations for their inclusive
Run 2 jet results.
Note that this algorithm specifies the transverse momentum $p_{T}$
rather than the transverse energy $E_{T}$.\footnote{%
The Run 2 jet algorithm results in cross section predictions
for CDF that are $\sim 7$\% lower than those using
the Run 1 algorithm, over most of the kinematic range.}

\begin{figure}[ht]
\begin{center}
\includegraphics[width=0.79\textwidth]{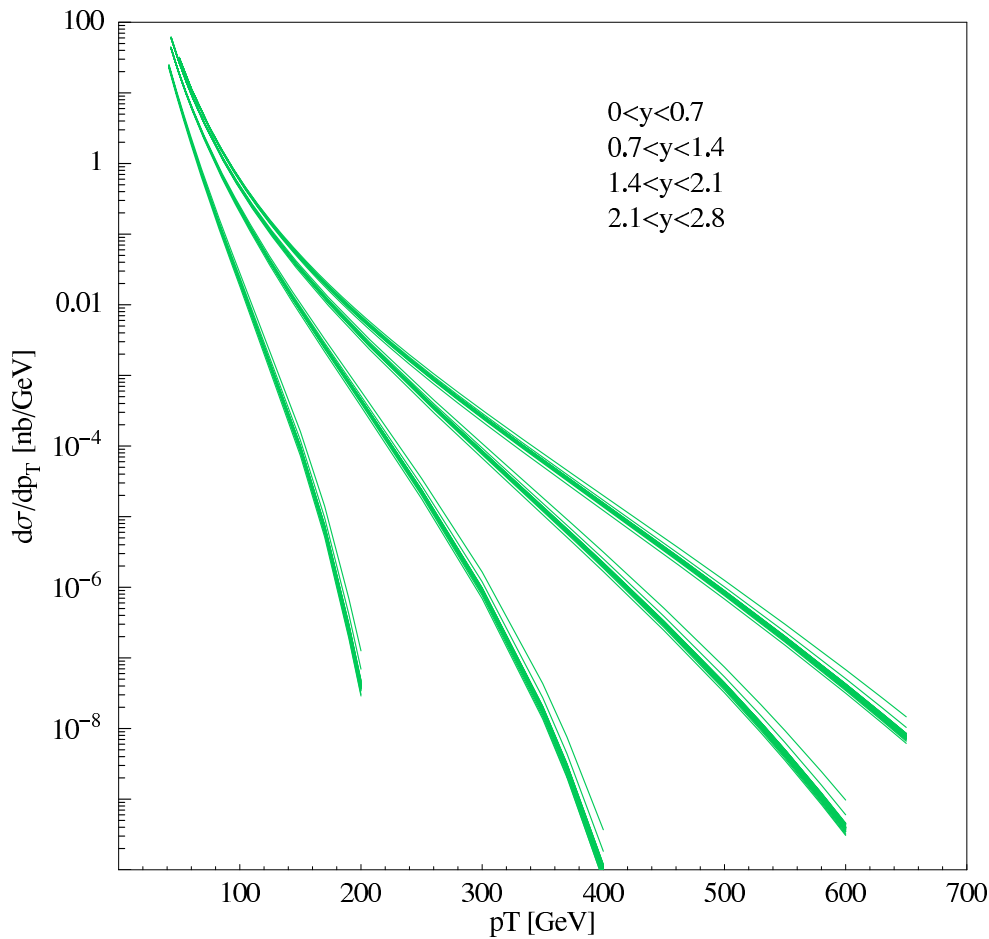}
\end{center}
\caption{Run 2 cross section for the CDF rapidity bins.
\label{fig:CDFRun2}}
\end{figure}

\begin{figure}[ht]
\begin{center}
\includegraphics[width=0.79\textwidth]{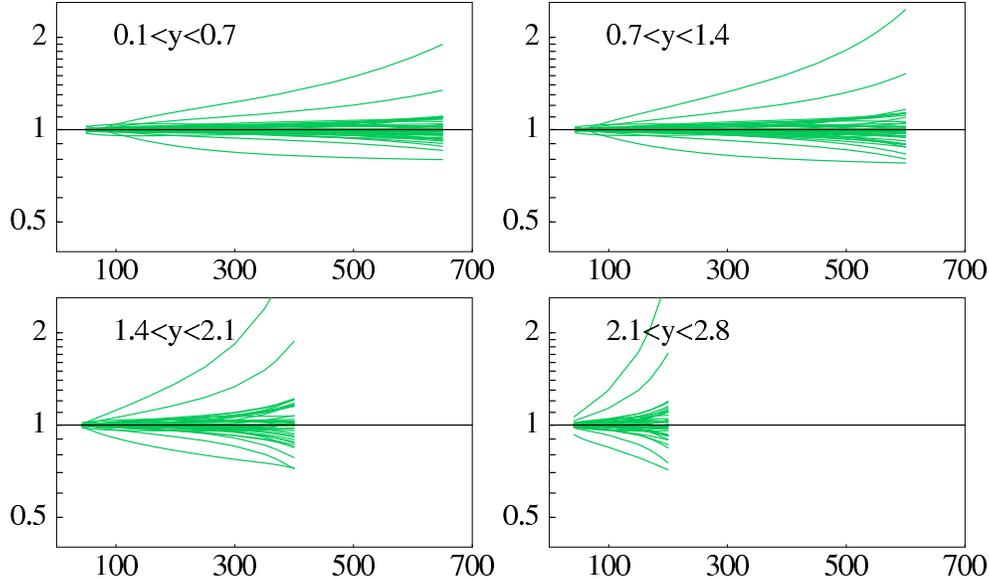}
\end{center}
\caption{Uncertainty range of the Run 2 cross section
for the CDF rapidity bins.
The curves show the ratios of the 40 eigenvector basis
sets compared to the central (CTEQ6.1M) prediction (ordinate)
versus $p_{T}$ in GeV (abscissa).
\label{fig:CDFRun2ratio}}
\end{figure}

For completeness we also show the plots of the
predictions of the inclusive jet cross section
at the Tevatron Run 2 for the CDF choice of
rapidity bins.
Figure \ref{fig:CDFRun2} shows the cross section
as a function of $p_{T}$ on a log scale.
Figure \ref{fig:CDFRun2ratio} shows the uncertainty
band in the form of the ratios of $d\sigma/dp_{T}$
for the 40 eigenvector basis sets compared to
the central prediction.

While the statistical errors in Runs 2a and 2b will be small,
so that the jet cross section can be extended to high $p_{T}$,
the experience from Run 1b shows that the systematic errors
will have a dominant effect on the significance of the results.
However, we can anticipate that the PDF's,
especially the gluon distribution, will be pinned down
at large $x$ by these data.
The experimental collaborations will make the data
most useful for global analysis by providing detailed
information on systematic uncertainties.

\subsection{The ratio of Run 2 to Run 1b cross sections}

\begin{figure}[ht]
\begin{center}
\includegraphics[width=0.79\textwidth]{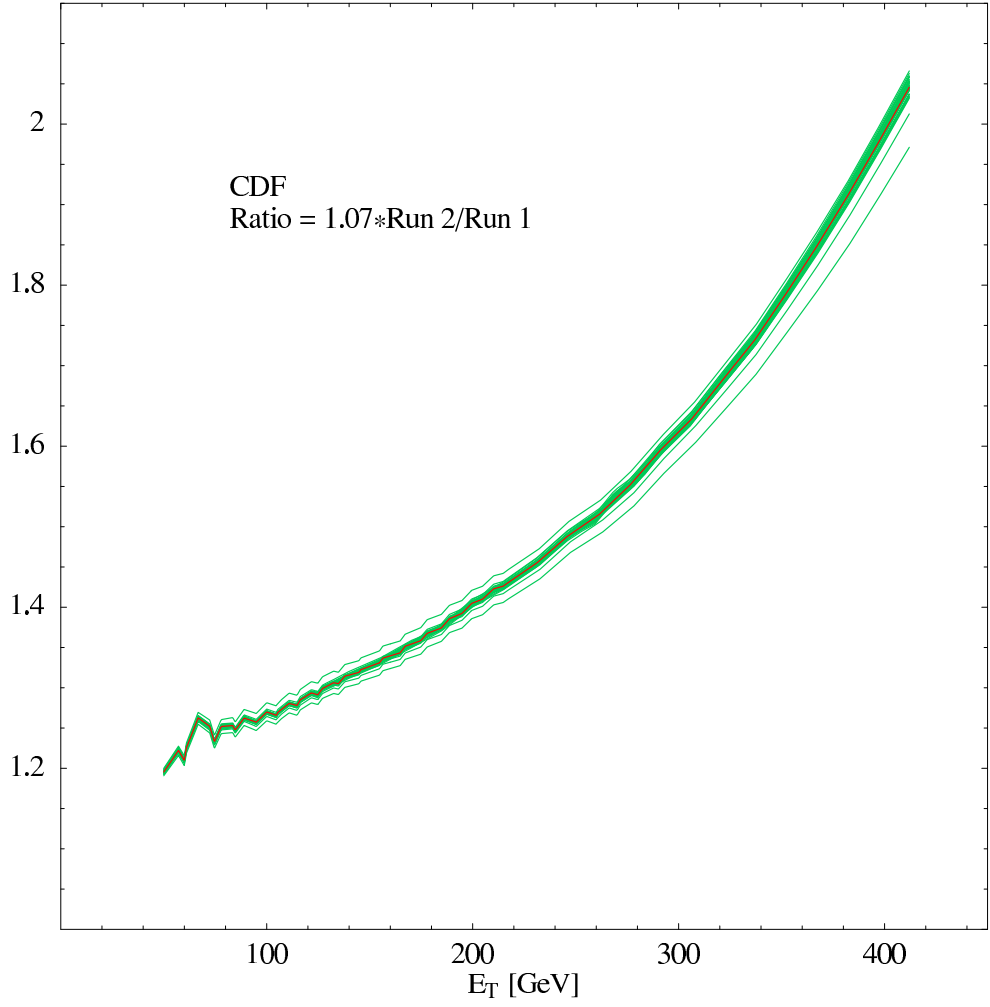}
\end{center}
\caption{The ratio of the Run 2 to Run 1b cross sections for the
central rapidity range ($0.1<|\eta|<0.7$).
The Run 2 cross section has been multiplied by 1.07
to take into account the difference in the jet
definitions for the Run 2 and Run 1b analyses.
(The small fluctuations are due to the Monte Carlo errors
in the calculations of the K-factors (NLO/LO) at each energy.)
\label{fig:CDFRatio2to1}}
\end{figure}

As mentioned previously, the Run 2 center-of-mass energy
$\sqrt{s}$ is higher than that from Run 1b.
The increase in $\sqrt{s}$ will have only a modest
impact on most physics cross sections.
However, it will lead to a rather large increase
in the production of high-$E_{T}$ jets.
It may be useful to examine the ratio of jet production
(for the same $E_{T}$ values) at the two different energies.
In such a ratio, many of the theoretical (and experimental)
errors will cancel.
The Run 2 to Run 1b ratio for the 40 sets of PDF's is shown
in Fig.\ \ref{fig:CDFRatio2to1} for the central rapidity
range of CDF ($0.1 < |\eta| < 0.7$).
As can be observed the ratio has a rather narrow theoretical
error band from PDF uncertainties, i.e., smaller uncertainty
than that of the absolute prediction of the
inclusive cross section.\footnote{%
Here we have multiplied our Run 2 predictions by a
factor of $1.07$ so that the comparison effectively
uses the Run 1 algorithm for both the numerator and denominator.
This is done for convenience since CDF will present its first
results for the Run 2 jet cross section using the Run 1
algorithm.}

\clearpage
\setcounter{section}{5}

\section{Inclusive jet production at the LHC}

\begin{figure}[ht]
\begin{center}
\includegraphics[scale=1]{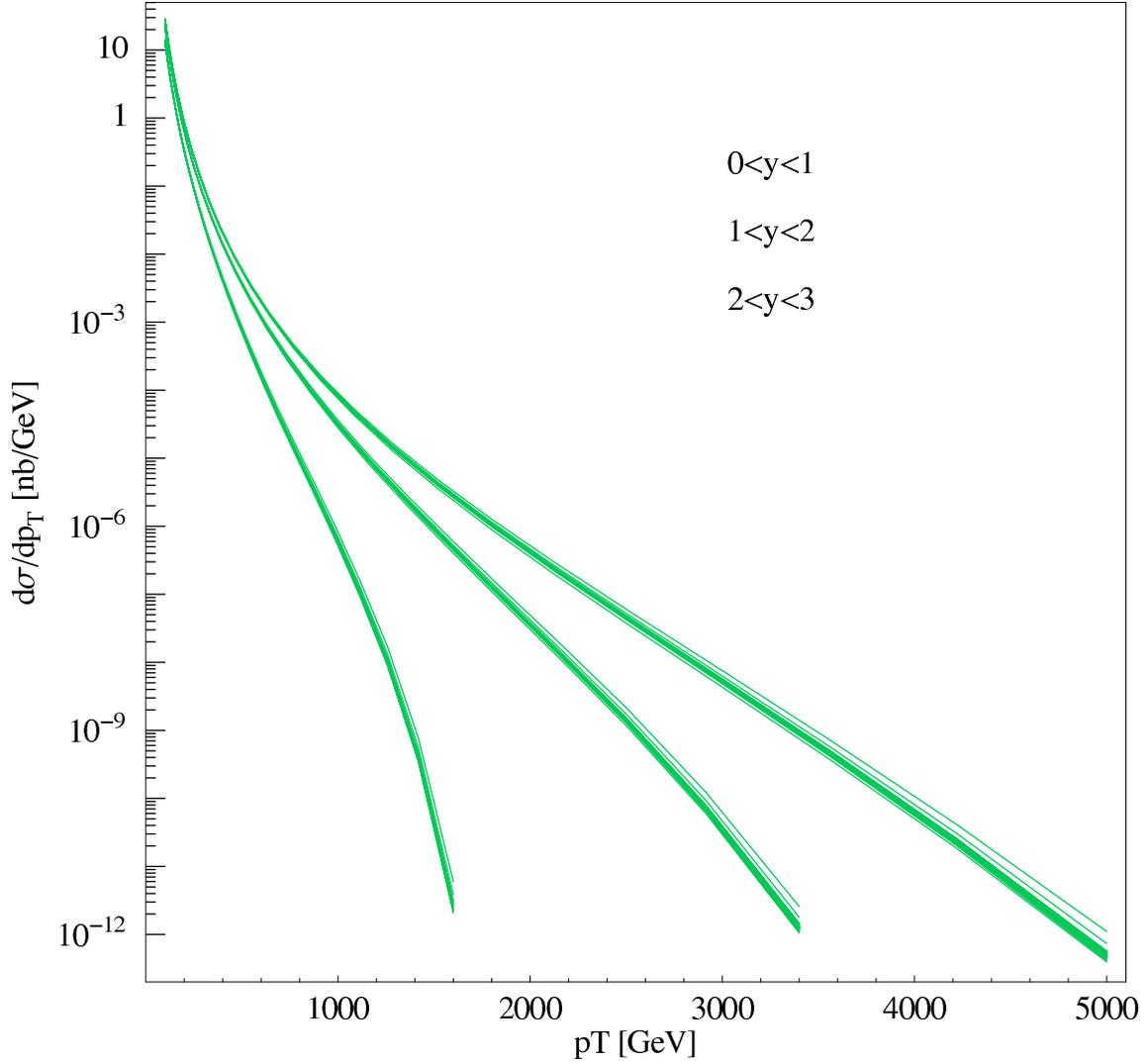}
\end{center}
\caption{The inclusive jet cross section as a function
of $p_{T}$ for three rapidity bins at the LHC.
The three rapidity ranges are $(0,1)$, $(1,2)$ and $(2,3)$.
Predictions of all 40 eigenvector basis sets are superimposed.
\label{fig:LHCcs}}
\end{figure}

\begin{figure}[ht]
\begin{center}
\includegraphics[scale=1]{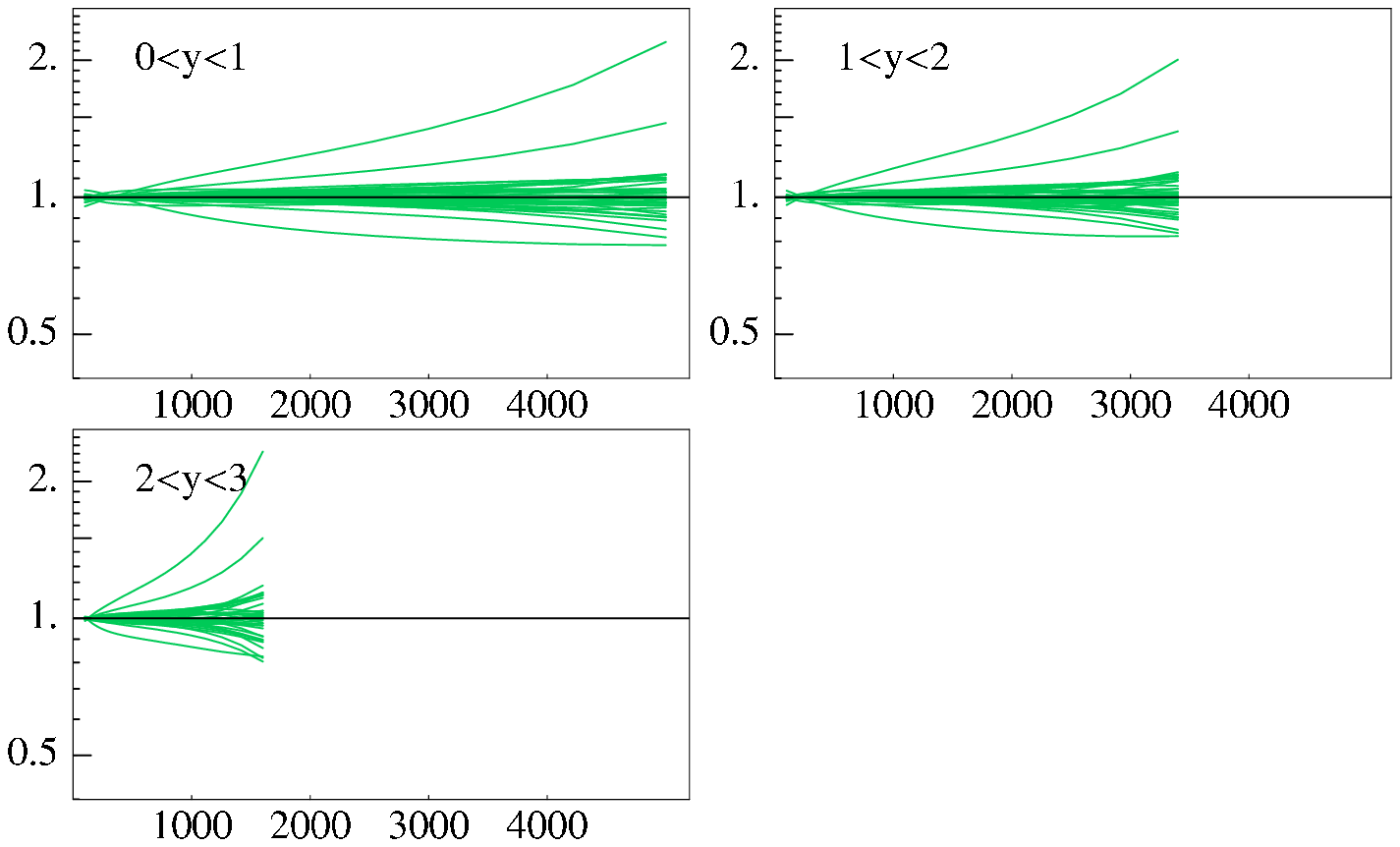}
\end{center}
\caption{The uncertainty range of the inclusive jet cross
section at the LHC.
The curves are graphs of the ratios of the cross sections
for the 40 eigenvector basis sets compared to the central
(CTEQ6.1M) prediction (ordinate) versus $p_{T}$ in GeV
(ordinate).
\label{fig:LHCrat}}
\end{figure}

The increase of the center-of-mass energy to $14$\,TeV
at the LHC will result in a dramatically larger accessible
kinematic range.
Inclusive jet cross sections will be measured out to transverse
momentum of $5$\,TeV/$c$ in the central rapidity range and out
to $1.5$\,TeV/$c$ in the forward rapidity region.
Figure \ref{fig:LHCcs} shows the predictions with uncertainty
ranges for three rapidity intervals.
Figure \ref{fig:LHCrat} shows the ratios of the
40 eigenvector basis set predictions to the central prediction.
The cross section uncertainties near the kinematic limit
at the LHC due to PDF uncertainties are similar in magnitude
to those obtained for Run 2 at the Tevatron.
Again the extremes of the predictions are provided by
the eigenvector basis sets $+15$ and $-15$, which correspond to 
extremes of the gluon distribution at large $x$.

The jet cross section at the LHC will be a discovery mode
for new physics.
Supersymmetry would produce an enhanced cross section
from production of gluino and squark jets.
Extra dimensions would show up from production of
Kaluza-Klein modes of the graviton.
So the pure QCD prediction will be compared to
the data and any difference will be the first evidence
for these or other new physics scenarios.

\clearpage
\setcounter{section}{6}
\section{Conclusions}

Jet production serves as a probe of the highest $Q^2$ scales accessible
at a hadron-hadron collider.  NLO QCD provides a good description of the
inclusive jet cross sections measured by CDF and D0 in Run 1 at the
Tevatron, provided that the gluon distribution is enhanced at high $x$ 
relative to earlier estimates.
Such an outcome is now the standard consequence of the global fitting
procedure due to the statistical power of the Tevatron jet data, and the
preference for a larger gluon in all rapidity regions. The uncertainties
on the jet cross section predictions still remain large, however,
primarily due to the remaining uncertainties on the gluon distribution.
In most regions, they are comparable to the experimental systematic
uncertainties.

The standard choice for the renormalization and factorization scales in
evaluating inclusive jet cross sections has been $E^{jet}_T/2$. We have
re-done the global fitting using alternate scales of $E^{jet}_T$ and
$2E^{jet}_T$ and have found changes to the gluon distribution that are
well within the gluon PDF uncertainty band. The effects of the
resummation of threshold logarithms has been calculated for the first
time in the forward rapidity regions. As for the central rapidity
region, the effects of the threshold logarithms are relatively small.

The increase of the center-of-mass energy to 1.96 TeV (from 1.8 TeV) and
the expected greater integrated luminosity will lead to an expanded
kinematic range for Run 2 at the Tevatron. Predictions have been made
for both the inclusive jet cross section and its uncertainty for Run 2
and for the LHC. The PDF uncertainty on the jet cross section near the
limits of the data reach in both cases are similar to those obtained for
Run 1.

We have carried out an exercise where we have examined the allowed
range for any new physics such as compositeness, given the current
scale of uncertainties on the jet cross section predictions.
Given the Run 1b jet data, any new physics must have
a scale larger than 1.6 TeV.
When data is available from Run 2 at the Tevatron,
and from the LHC, analogous analyses will be used to
identify evidence for new physics or place even
stricter limits on new physics.

\clearpage

\appendix{\noindent \bf Appendix}
\label{sec:scale}

One component of the theoretical uncertainty of any perturbative QCD 
calculation is the dependence of the answer on the various renormalization 
and factorization scales. One virtue of using NLO calculations in global 
determinations of parton distributions is the reduction in scale dependence 
usually exhibited by such calculations. The following 
brief discussion outlines the origin of the reduction of the scale dependence 
which is often observed with NLO calculations. 

Consider a large transverse momentum process such as the single jet inclusive 
cross section involving only massless partons. Furthermore, in order to 
simplify the notation, suppose that the transverse momentum is sufficiently 
large that only the quark distributions need be considered. In the following,  
a sum over quark flavors is implied. Schematically, one can write the 
lowest order cross section as 
\begin{equation}
E{d^3\sigma\over dp^3}\equiv \sigma = a^2(\mu)\, \hat \sigma_B\otimes q(M)
\otimes q(M)
\end{equation}

\noindent where $a(\mu)=\alpha_s(\mu)/2\pi$ and the lowest order 
parton-parton scattering cross section is denoted by $\hat \sigma_B$.
The renormalization and factorization scales are denoted by $\mu {\ and\ } 
M$, 
respectively. In addition, various overall factors have been absorbed into the 
definition of $\hat \sigma_B$.
The symbol $\otimes$ denotes a convolution defined as 
\begin{equation}
f\otimes g = \int_x^1 {dy \over y}\, f({x \over y}) \, g(y).
\end{equation}

When one calculates the ${\cal O}(\alpha_s^3)$ contributions to the inclusive 
cross section, the result can be written as 
\begin{eqnarray}
\sigma & = & a^2(\mu)\,  \hat \sigma_B\otimes q(M)\otimes q(M) \nonumber \\
& + & 2 a^3(\mu)\,  b \ln(\mu/p_T) \hat \sigma_B\otimes q(M)\otimes q(M) 
\nonumber \\
& + & 2 a^3(\mu)\, \ln(p_T/M) P_{qq}\otimes \hat \sigma_B \otimes q(M)\otimes 
q(M) \nonumber \\
& + & a^3(\mu)\, K \otimes q(M) \otimes q(M).
\label{eqn:sigma}
\end{eqnarray}

In writing Eq.\ (\ref{eqn:sigma}), specific logarithms associated with the 
running coupling 
and the scale dependence of the parton distributions have been explicitly 
displayed; the remaining higher order corrections have been collected in the 
function $K$ in the last line of Eq.\ (\ref{eqn:sigma}). The $\mu$ dependence 
of the running coupling is obtained from 
\begin{equation}
\mu {\partial a(\mu)\over \partial \mu} = \beta(a(\mu))
\label{eqn:beta}
\end{equation}
where $\beta = -b a^2 (1+c a)$ with 
\begin{equation}
b={33-2 f\over 6} {\rm \ and \ } c={153 - 19 f\over 2(33-2 f)}
\end{equation}
where $f$ denotes the number of flavors. The scale dependence 
of the quark distributions is given by the nonsinglet DGLAP
equation~\cite{DGLAP}
\begin{equation}
M {\partial q(x,M)\over \partial M} = a(M) P_{qq}\otimes q(M).
\end{equation}

Now, the NLO expression in Eq.\ (\ref{eqn:sigma})
should be independent of the choice of scale up to corrections
of order $a^4$.
That this is indeed the case is easy to see.
For convenience, set $M=\mu$ and calculate the logarithmic 
derivative of $\sigma$ with respect to $\mu$, 
\begin{eqnarray}
\mu {\partial \sigma \over \partial \mu} = & 
2 a(\mu) [-b a^2(\mu)(1+c a(\mu))] \hat \sigma \otimes q(\mu) \otimes q(\mu) 
\cr & +2 a^3(\mu) P_{qq} \otimes \hat \sigma_B \otimes q(\mu) \otimes q(\mu) 
\nonumber \\
& + 2 b a^3(\mu) \hat \sigma_B \otimes q(\mu) \otimes q(\mu) \nonumber \\ 
& - 2 a^3(\mu) P_{qq}\otimes \hat \sigma_B \otimes q(\mu) \otimes q(\mu) 
+ \cdots \nonumber \\
& = 0 + {\cal O}(a^4).
\label{eqn:deriv}
\end{eqnarray}
The derivative of the lowest order contribution yields the first two
lines of Eq.\ (\ref{eqn:deriv}) while the derivative of the logarithms
in the second and third lines of Eq.\ (\ref{eqn:sigma}) 
give rise to the last two lines in Eq.\ (\ref{eqn:deriv}) which cancel 
the $\mu$ dependent terms coming from the derivative of the lowest-order
contribution.
All other derivatives yield contributions which are of order $a^4$.
Thus, the structure of the NLO calculation is such that the scale
dependence is cancelled up to terms of order $a^4$.

The preceding discussion shows that one might expect a reduced scale 
dependence for the NLO calculation as opposed to that observed at lowest 
order.
However, the uncancelled ${\cal O}(a^4)$ and higher terms will still 
result in some scale dependence.
It is easy to understand the systematics of 
the remaining dependence for the case at hand.
Consider, first, the lowest-order contribution shown in the
first line of Eq.\ (\ref{eqn:sigma}).
At large values of $p_T$ as $x_T =2 p_T/\sqrt s$ tends to one,
the scaling violations of the parton distributions lead to a
decrease in the cross section as the scale increases.
In addition, the running coupling decreases as the scale is increased.
Therefore, at fixed $p_T$ the lowest order contribution exhibits 
a monotonically decreasing behavior as the scale increases. 

The second line of Eq.\ (\ref{eqn:sigma}) contains the contribution
which partially offsets the scale dependence of the running coupling.
Apart from the explicit logarithm, this contribution has much the
same scale dependence as the lowest order contribution.
Thus, for $\mu < p_T$ this term is negative, becoming 
positive as the scale is increased beyond $p_T$. 

The third line contains the contribution which partially offsets the 
scale dependence originating in the parton distributions.
In the kinematic region under consideration,
the convolution of the splitting function $P_{qq}$ with the
lowest contribution gives a negative result, reflecting 
the fact that the parton distributions decrease with increasing scale at 
large values of $x$.
Therefore, this contribution is negative when the 
factorization scale $M$ is chosen to be less than $p_T$. 

Finally, the remaining term on line four of Eq.\ (\ref{eqn:sigma})
has much the same scale dependence as the lowest order term,
as the dominant scale dependence has already been exhibited
in the terms on lines two and three.
Thus, lines one and four give contributions which decrease
monotonically with increasing scale while lines two and three
start out negative, reach zero when the scales are equal to
$p_T$, and are positive for larger scales.
This behavior explains why the typical NLO scale dependence
in the high-$p_T$ region yields a result which first increases
with increasing scale, reaches a maximum, and then decreases.
For many high-$p_T$ processes involving massless partons this
maximum occurs near $\mu = M = p_T$. 

Next, it is interesting to consider how the location of the maximum changes 
with $p_T$. Suppose that one goes to small values of $x_T$ such that the 
typical values of $x$ in the parton distributions is around .2 or smaller. 
In this region the scaling violations change sign, corresponding to a growth 
of the parton distributions with increasing scale, rather than a decrease. 
The contribution of the third line of Eq.(1) then starts to compete against 
that of the second line. Accordingly, one must go to smaller values of 
$\mu/p_T$ before seeing the turnover and decrease with decreasing scale. That 
is, the position of the maximum at fixed $p_T$ moves to lower values of 
$\mu/p_T$ as $p_T$ decreases. The converse is also clearly true; as $p_T$ 
increases the position of the maximum moves to higher values of $\mu/p_T$. 
Thus, if one were to try to find an optimized scale at each value of $p_T$, 
the pattern that would emerge would be that the optimal scale would be a 
fractional multiple of $p_T$ at lower $p_T$ values and would increase to 
a multiple of $p_T$ greater than one at higher $p_T$ values. This pattern 
is precisely what is observed when using the single scale version of the 
Principle of Minimal Sensitivity \cite{PMS}.

Another manifestation of this pattern concerns the behavior of the 
``K factor'', defined as the ratio of the NLO to LO result.
At a fixed value of $p_T$, the K factor will first rise and then fall with 
increasing scale. Suppose that at some $p_T$ value the maximum occurs 
for the choice $\mu = M = p_T/2$. Then, as one goes to higher $p_T$ values 
the previous discussion shows that the maximum will move to a higher scale 
choice such as $p_T$ or even $2 p_T$. At the larger $p_T$ values it will then 
appear that the K factor may increase with increasing scale. This is just the 
behavior seen in the calculations done for the high rapidity D0 jet data.


\end{document}